\documentclass[sigconf]{acmart}

\setcopyright{none} 
\acmYear{2020}

\usepackage{tikz}
\usepackage{amsmath}
\usepackage{multirow}
\usepackage{makecell}
\usepackage{verbatim}
\usepackage{subfig}
\usepackage{url}

\usepackage{markdown}

\usepackage{ulem}
\usepackage{pifont}

\usepackage{graphicx}
\usepackage{color, colortbl}

\usepackage{listings}
\usetikzlibrary{positioning,shadows.blur,fit}
\definecolor{auburn}{rgb}{0.43, 0.21, 0.1}
\definecolor{codegreen}{rgb}{0,0.6,0}
\definecolor{codegray}{rgb}{0.5,0.5,0.5}
\definecolor{codepurple}{rgb}{0.58,0,0.82}
\lstdefinelanguage
   [x64]{Assembler}     
   [x86masm]{Assembler} 
   {morekeywords={CDQE,CQO,CMPSQ,CMPXCHG16B,JRCXZ,LODSQ,MOVSXD,movl,addl,subl, %
                  pushq,popq,leaq,cmpq,POPFQ,PUSHFQ,SCASQ,STOSQ,IRETQ,RDTSCP,SWAPGS, %
                  rax,rdx,rcx,rbx,rsi,rdi,rsp,rbp,retq,callq,movq, %
                  r8,r8d,r8w,r8b,r9,r9d,r9w,r9b,r10,r11,r12,r15,enclu,mfence,lfence,cmova,cmovg,cmovl}} 
\lstdefinestyle{mystyle}{
    backgroundcolor=\color{white},   
    commentstyle=\color{blue},
    keywordstyle=\color{auburn},
    numberstyle=\tiny\color{codegray},
    stringstyle=\color{codepurple},
    basicstyle=\footnotesize,
    breakatwhitespace=false,         
    breaklines=true,                 
    captionpos=b,                    
    keepspaces=true,                 
    numbers=left,                    
    numbersep=5pt,                  
    showspaces=false,                
    showstringspaces=false,
    showtabs=false,                  
    tabsize=2
}
\usepackage{amsmath,amssymb,amsfonts}
\usepackage{algorithmic}
\usepackage{graphicx}
\usepackage{textcomp}
\usepackage{xcolor}
\def\BibTeX{{\rm B\kern-.05em{\sc i\kern-.025em b}\kern-.08em
    T\kern-.1667em\lower.7ex\hbox{E}\kern-.125emX}}
    
\usepackage{appendix}

\newcommand{\ignore}[1]{}

\newcommand\wenhao[1]{\textcolor{magenta}{\{\textbf{wenhao:} {\em#1}\}}}

\newcommand\weijie[1]{\textcolor{brown}{\{\textbf{weijie:} {\em#1}\}}}

\begin{document}

\title{Confidential Attestation: Efficient in-Enclave Verification of Privacy Policy Compliance}

\author{Weijie Liu}
\email{weijliu@iu.edu}
\affiliation{%
  \institution{Indiana University Bloomington}
  \country{USA}
}

\author{Wenhao Wang}
\email{wangwenhao@iie.ac.cn}
\affiliation{%
  \institution{Institute of Information Engineering, CAS}
  \country{China}
}

\author{XiaoFeng Wang}
\email{xw7@indiana.edu}
\affiliation{%
  \institution{Indiana University Bloomington}
  \country{USA}
}

\author{Xiaozhu Meng}
\email{xm13@rice.edu}
\affiliation{%
 \institution{Rice University}
 \country{USA}
}

\author{Yaosong Lu}
\email{luyaosong@iie.ac.cn}
\affiliation{%
  \institution{SKLOIS, Institute of Information Engineering, CAS}
  \country{China}
}

\author{Hongbo Chen}
\email{hc50@iu.edu}
\affiliation{%
  \institution{Indiana University Bloomington}
  \country{USA}
}

\author{Xinyu Wang}
\email{wang_x_y@sjtu.edu.cn}
\affiliation{
  \institution{Shanghai Jiao Tong University}
  \country{China}
}

\author{Qintao Shen}
\email{shenqintao@iie.ac.cn}
\affiliation{%
  \institution{SKLOIS, Institute of Information Engineering, CAS}
  \institution{School of Cyber Security, University of Chinese Academy of Sciences}
  \country{China}
}

\author{Kai Chen}
\email{chenkai@iie.ac.cn}
\affiliation{%
  \institution{Institute of Information Engineering, CAS}
  \country{China}
}

\author{Haixu Tang}
\email{hatang@indiana.edu}
\affiliation{%
  \institution{Indiana University Bloomington}
  \country{USA}
}

\author{Yi Chen}
\email{chenyi@iie.ac.cn}
\affiliation{%
  \institution{SKLOIS, Institute of Information Engineering, CAS}
  \country{China}
}

\author{Luyi Xing}
\email{luyixing@indiana.edu}
\affiliation{%
  \institution{Indiana University Bloomington}
  \country{USA}
}

\begin{abstract}
A trusted execution environment (TEE) such as Intel Software Guard Extension (SGX) runs a remote attestation to prove to a data owner the integrity of the initial state of an enclave, including the program to operate on her data. For this purpose, the data-processing program is supposed to be open to the owner or a trusted third party, so its functionality can be evaluated before trust can be established. In the real world, however, increasingly there are application scenarios in which the program itself needs to be protected (e.g., proprietary algorithm). So its compliance with privacy policies as expected by the data owner should be verified without exposing its code.

To this end, this paper presents CAT, a new model for TEE-based confidential attestation. Our model is inspired by Proof-Carrying Code (PCC), where a code generator produces proof together with the code and a code consumer verifies the proof against the code on its compliance with security policies. Given that the conventional solutions do not work well under the resource-limited and TCB-frugal TEE, we come up with a new design that allows an untrusted out-enclave generator to analyze and instrument the source code of a program when compiling it into binary and a trusted in-enclave consumer efficiently verifies the correctness of the instrumentation and the presence of other protection before running the binary. Our design strategically moves most of the workload to the code generator, which is responsible for producing well-formatted and easy-to-check code, while keeping the consumer simple. Also, the whole consumer can be made public and verified through a conventional attestation. We implemented this model on Intel SGX and demonstrate that it introduces a very small part of TCB (2000 LoCs), compared with the sandbox solution or theorem-proving based PCC. We also thoroughly evaluated its performance on micro- and macro- benchmarks and real-world applications, showing that the new design only incurs a small overhead (no more than 23\%) when enforcing several categories of security policies.


\end{abstract}





\maketitle

\section{Introduction}\label{sec-introduction}


Recent years have witnessed the emergence of hardware trusted execution environments (TEEs) that enable efficient computation on untrusted platforms. A prominent example is Intel SGX~\cite{mckeen2013innovative}, a TEE widely deployed on commercial-off-the-shelf (COTS) desktops and server processors, providing secure memory called \textit{enclave} to host confidential computing on sensitive data, which are protected from the adversary in control of the operating system and even with physical access to the data-processing machine. Such a computing model has already been supported by major cloud providers today, including Microsoft Azure and Google Cloud~\cite{russinovich2017introducing,asylo2019}, and its further adoption has been facilitated by the Confidential Computing Consortium~\cite{ccc2019}, a Linux Foundation project that brings together the biggest technical companies such as Intel, Google, Microsoft and IBM etc. However, before TEEs can see truly wide deployment for real-world confidential computing, key technical barriers still need to be overcome, \textit{remote attestation} in particular.


\vspace{3pt}\noindent\textbf{Remote attestation}. At the center of a TEE's trust model is remote attestation (RA), which allows the user of confidential computing to verify that the enclave code processing her sensitive data is correctly built and operates on a genuine TEE platform, so her data is well protected. This is done on SGX through establishing a chain of trust rooted at a platform attestation key owned by the hardware manufacturer and using the key to generate a \textit{Quote} -- a signed report that contains the measurement of the code and data in an enclave; the Quote is delivered to the data owner and checked against the signature and the expected measurement hash. This trust building process is contingent upon the availability of the measurement, which is calculated from the enclave program either by the data owner when the program is publicly available or by a trusted third party working on the owner's behalf. This becomes problematic when the program itself is private and cannot be exposed.
Programs may have exploitable bugs or they may write information out of the enclave through corrupted pointers easily.
For example, different banks and financial agencies would like to jointly calculate a person's credit score based on each other's data, without disclosing their individual data and the proprietary algorithm processing it. As another example, the pharmaceutical companies want to search for suitable candidates for their drug trial without directly getting access to plaintext patient records or exposing their algorithm (carrying sensitive genetic markers discovered with million dollar investments) to the hospital. With applications of this kind on the rise, new techniques for protecting both data and code privacy are in great demand.


\vspace{3pt}\noindent\textbf{Confidential attestation: challenges}. To address this problem, we present in this paper a novel \textit{Confidential ATtestation} (\textit{CAT}) model to enable verification of an enclave program's compliance with user-defined security policies without exposing its source or binary code to unauthorized parties involved. Under the CAT model, a \textit{bootstrap enclave} whose code is public and verifiable through the Intel's remote attestation, is responsible for performing the compliance check on behalf of the participating parties, who even without access to the code or data to be attested, can be convinced that desired policies are faithfully enforced. 
However, building a system to support the CAT model turns out to be nontrivial, due to the complexity in static analysis of enclave binary for policy compliance, the need to keep the verification mechanism, which is inside the enclave's \textit{trusted computing base} (\textit{TCB}), small, the demand for a quick turnaround from the enclave user, and the limited computing resources today's SGX provides (about 96 MB physical memory on most commercial hardware~\cite{chakrabarti2019scaling}). 
Simply sand-boxing the enclave code significantly increases the size of TCB, rendering it less trustworthy, and also brings in performance overheads incurred by confinement and checkpoint/rollback~\cite{hunt2018ryoan}.

A promising direction we envision that could lead to a practical solution is \textit{proof-carry code} (\textit{PCC}), a technique that enables a \textit{verification condition generator} (\textit{VCGen})~\cite{colby2000certifying,leroy2006formal,pirzadeh2010extended} to analyze a program and create a proof that attests the program's adherence to policies, and a \textit{proof checker} to verify the proof and the code. The hope is to push the heavy-lifting part of the program analysis to the VCGen outside the enclave while keeping the proof checker inside the enclave small and efficient.  The problem is that this \textit{cannot} be achieved by existing approaches, which utilize formal verification (such as~\cite{necula2001oracle,pirzadeh2010extended}) to produce a proof that could be 1000$\times$ larger than the original code. Actually, with years of development, today's formal verification techniques, theorem proving in particular, are still less scalable, unable to handle large code blocks (e.g., over 10000 instructions) when constructing a security proof.

\vspace{3pt}\noindent\textbf{Our solution}. In our research, we developed a new technique to instantiate the CAT model on SGX. Our approach, called \textit{CAT-SGX}, is a PCC-inspired solution, which relies on out-of-enclave targeted instrumentation for lightweight in-enclave information-flow confinement and integrity protection, instead of heavyweight theorem proving. More specifically, CAT-SGX operates an untrusted \textit{code producer} as a compiler to build the binary code for a data-processing program (called \textit{target program}) and instrument it with a set of \textit{security annotations} for enforcing desired policies at runtime, together with a lightweight trusted \textit{code consumer} running in the bootstrap enclave to statically verify whether the target code indeed carries properly implanted security annotations.



To reduce the TCB and in-enclave computation, CAT-SGX is designed to simplify the verification step by pushing out most computing burden to the code producer running outside the enclave. More specifically, the target binary is expected to be well formatted by the  producer, with all its indirect control flows resolved, all possible jump target addresses specified on a list and enforced by security annotations.  In this way, the code consumer can check the target binary's policy compliance through lightweight \textit{Recursive Descent Disassembly} to inspect its complete control flow (Section~\ref{subsec-boundarychecking}), so as to ensure the presence of correctly constructed security annotations in front of each critical operation, such as read, store, enclave operations like OCall, and stack management (through a shadow stack). Any failure in such an inspection causes the rejection of the program. Also, since most code instrumentation (for injecting security annotations) is tasked to the producer, the code consumer does not need to make any change to the binary except relocating it inside the enclave. As a result, we only need a vastly simplified disassembler, instead of a full-fledged, complicated binary analysis toolkit, to support categories of security policies, including data leak control, control-transfer management, self-modifying code block and side/covert channel mitigation
(Section~\ref{subsec-policies}).  A wider spectrum of policies can also be upheld by an extension of CAT-SGX, as discussed in the paper (Section~\ref{sec-discussion}).

We implemented CAT-SGX in our research, building the code producer on top of the LLVM compiler infrastructure and the code consumer based upon the Capstone disassembly framework~\cite{capstone} and the core disassembling engine for X86 architecture. 
Using this unbalanced design, our in-enclave program has only 2000 lines of source code, and together with all the libraries involved, it is compiled into 1.9 MB binary. This is significantly smaller than the NaCl's core library used by Ryoan, whose binary is around 19 MB and the theorem prover Z3, with 26 MB. We further evaluated our implementation on micro-benchmarks (nBench), as well as macro-benchmarks, including credit scoring, HTTPS server, and also basic biomedical analysis algorithms (sequence alignment, sequence generation, etc.) over various sizes of genomic data (1000 Genomes Project~\cite{1000genomes}), under the scenario of confidential computing as a service (Section~\ref{subsec-scenarios}). 
CAT-SGX incurs on average (calculated by geometric mean) 20\% performance overhead and less than 30\% storage overhead enforcing all the proposed security policies, and leads to around 10\% performance overhead and less than 20\% storage overhead without side/covert channel mitigation.
We have released our code on Github~\cite{our-prototype}.

\vspace{3pt}\noindent\textbf{Contributions}. The contributions of the paper are outlined as follows:

\vspace{3pt}\noindent$\bullet$\textit{ Confidential attestation model}. We propose CAT, a new model that extends today's TEE to maintain the data owner's trust in protection of her enclave data, without exposing the code of the data-processing program. This is achieved through enforcing a set of security policies through a publicly verifiable bootstrap enclave. This new attestation model enables a wide spectrum of applications with great real-world demand in the confidential computing era. 
    
\vspace{3pt}\noindent$\bullet$\textit{ New techniques for supporting CAT on SGX}. We present the design for instantiating CAT on SGX, following the idea of PCC. Our approach utilizes out-of-enclave code analysis and instrumentation to minimize the workload for in-enclave policy compliance check, which just involves a quick run of a well-formatted target binary for inspecting correct instrumentation. This simple design offers supports for critical policies, ranging from memory leak prevention to side channel mitigation\ignore{ to multi-thread isolation}, through a much smaller TCB compared with sand-box solutions.


\vspace{3pt}\noindent$\bullet$\textit{ Implementation and evaluation}. We implemented our design of CAT-SGX and extensively evaluated our prototype on micro- and macro- benchmarks, together with popular biomedical algorithms on human genomic data.  Our experiments show that CAT-SGX effectively enforces various security policies at small cost, with the delay incurred by memory leak prevention around 20\% and side-channel mitigation usually no more than 35\%.

\section{Background}\label{sec-background}

\noindent\textbf{Intel SGX}. Intel SGX~\cite{mckeen2013innovative} is a user-space TEE, which is characterized by flexible process-level isolation: a program component can get into an enclave mode and be protected by execution isolation, memory encryption and data sealing, against the threats from the untrusted OS and processes running in other enclaves. Such protection, however, comes with in-enclave resource constraints. Particularly, only 128 MB encryption protected memory (called Enclave Page Cache or EPC) is reserved for enclaves for each processor. Although the virtual memory support is available, it incurs significant overheads in paging. 

Another problem caused by SGX's flexibility design is a large attack surface. When an enclave program contains memory vulnerabilities, attacks can happen to compromise enclave's privacy protection. Prior research demonstrates that a Return-Oriented-Programming (ROP) attack can succeed in injecting malicious code inside an enclave, which can be launched by the OS, Hypervisor, or BIOS~\cite{lee2017hacking,biondo2018guard,schwarz2019practical}. 
Another security risk is side-channel leak~\cite{schwarz2017malware,lee2017inferring,gras2018translation}, caused by the thin software stack inside an enclave (for reducing TCB),  which often has to resort to the OS for resource management (e.g., paging, I/O control). Particularly, an OS-level adversary can perform a controlled side channel attack (e.g.,~\cite{xu2015controlled}). Also in the threat model is the physical adversary, such as a system administrator, who tries to gain unauthorized access to a TEE’s computing units to compromise its integrity or confidentiality.

\vspace{3pt}\noindent\textbf{SGX remote attestation}. As mentioned earlier, attestation allows a remote user to verify that the enclave is correctly constructed and run on a genuine SGX-enabled platform. In Intel’s attestation model, three parties are involved: (1) the Independent Software Vendor (ISV) who is registered to Intel as the enclave developer; (2) the Intel Attestation Service (IAS) hosted by Intel to help enclave verification,
and (3) the SGX-enabled platform that operates SGX enclaves. The attestation begins with the ISV sending an attestation request challenge, which can be generated by an enclave user or a data owner who wants to perform the attestation with the enclave to check its state. Upon recipient of the challenge, the enclave then generates a verification report including the enclave measurement, which can be verified by a quoting enclave (QE) through \textit{local attestation}. The QE signs the report using the attestation key and the generated \textit{quote} is forwarded to the Intel Attestation Service (IAS). 
The IAS then checks the quote and signs the verification result using Intel's private key. The ISV can then validate the attestation result based upon the signature and the enclave measurement.

\vspace{3pt}\noindent\textbf{PCC}. PCC is a software mechanism that allows a host system to verify an application's properties with a formal proof accompanying the application's executable code. Using PCC, the host system is expected to quickly check the validity of the proof, and compare the conclusions of the proof to its own security policies to determine whether the application is safe to run. 
Traditional PCC schemes tend to utilize formal verification for proof generation and validation. Techniques for this purpose includes verification condition generator/proof generator~\cite{homeier1995mechanically,colby2000certifying}, theorem prover/proof assistant~\cite{paulson2000isabelle,de2008z3,bertot2013interactive}, and proof checker/verifier~\cite{appel2003trustworthy}, which typically work on type-safe intermediate languages (IL) or higher level languages. A problem here is that up to our knowledge, no formal tool today can automatically transform a binary to IL for in-enclave verification. BAP~\cite{brumley2011bap} disassembles binaries and lifts x86 instructions to a formal format, but it does not have a runtime C/C++ library
for static linking, as required for an enclave program.

Moreover, the PCC architecture today relies on the correctness of the VCGen and the proof checker, so a direct application of PCC to confidential computing needs to include both in TCB. This is problematic due to their complicated designs and implementations, which are known to be error-prone~\cite{necula2001oracle}. Particularly, today's VCGens are built on interpreter/compiler or even virtual machine~\cite{leroy2006formal}, and therefore will lead to a huge TCB. Prior attempts~\cite{appel2001foundational} to move VCGen out of TCB are found to have serious performance impacts, due to the significantly increased proof size.  

Actually, the proof produced by formal verification is typically large, growing exponentially with the size of the program that needs certified~\cite{necula1997proof}. It is common to have a proof 1000 times larger than the code~\cite{pirzadeh2010extended}.\ignore{ The proof/certificate in PCC is a formal representation that can be encoded as e.g. LF term. Such proof terms include a lot of repetition which means it includes huge certificates. } Although techniques are there to reduce the proof size~\cite{appel2001foundational,pirzadeh2010extended},\ignore{ e.g. OPCC introduces a non-deterministic proof checker that makes the proof 30 times smaller. However, }they are complicated and increase the TCB size~\cite{appel2003trustworthy}. Therefore as far as we are aware, no existing PCC techniques can be directly applied to enable the CAT model on today's TEE.

\section{Confidential Attestation}\label{sec-CAT}


Consider an organization that provides data-processing services, such as image editing (Pixlr), tax preparation (TurboTax), personal health analysis (23andMe) and deep learning inference as a service.  To use the services, its customers need to upload their sensitive data, such as images, tax documents, and health data, to the hosts operated by the organization. To avoid exposing the data, the services run inside SGX enclaves and need to prove to the customers that they are only accessible to authorized service programs. However, the organization may not want to release the proprietary programs to protect its intellectual property. This problem cannot be addressed by today's TEE design.  
In this section, we present the \textit{Confidential ATtestation} (CAT) model to allow the data owner to verify that the enclave code satisfies predefined security policy requirements without undermining the privacy of the enclave code.





\begin{figure}[htbp]
\centerline{\includegraphics[scale=0.75]{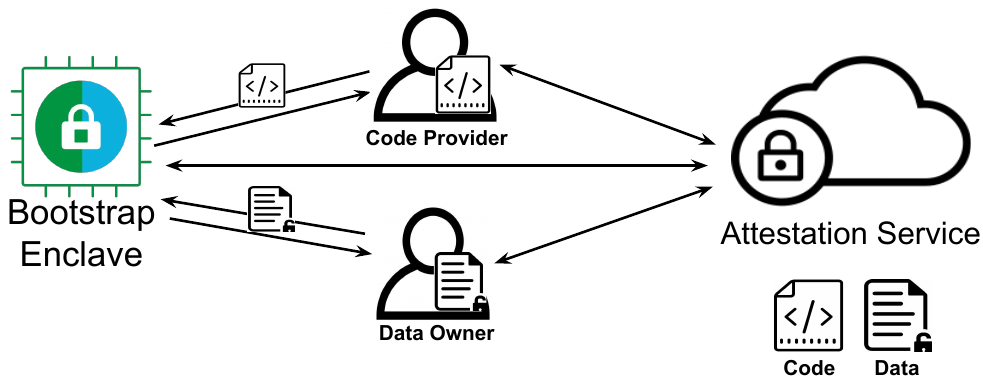}}
\caption{The CAT model}\label{fg-cat}
\end{figure}

\subsection{The CAT Model}
The CAT model can be described by the interactions among 4 parties, as follows: 

\vspace{3pt}\noindent\textbf{Attestation service}. Attestation service (AS) assists in the remote attestation process by helping the data owner and/or the code provider verify the quote generated by an enclave, as performed by the Intel attestation service for SGX. 


\vspace{3pt}\noindent\textbf{Bootstrap enclave}. The boostrap enclave is a built-in control layer on the software stack of an enclave supporting CAT (see Figure~\ref{fg-cat}). Its code is public and initial state is measured by hardware for generating an attestation quote, which is later verified by the data owner and the code provider with the help of the AS. This software layer is responsible for establishing security channels with enclave users, authenticating and dynamically loading the binary of the target program from the code provider and data from its owner. Further it verifies the code to ensure its compliance with predefined security policies before bootstrapping the computation. During the computing, it also controls the data entering or exiting the enclave, e.g., through SGX ECalls and OCalls to perform data sanitization.

\vspace{3pt}\noindent\textbf{Data owner}. The data owner uploads sensitive data (e.g., personal images) to use in-enclave services (e.g., an image classifier) and intends to keep her data secret during the computation. To this end, the owner runs a remote attestation with the enclave to verify the code of the bootstrap enclave, and sends in data through a secure channel only when convinced that the enclave is in the right state so expected policy compliance check will be properly performed on the target binary from the code provider.  Note that there could be more than one data owner to provide data.


\vspace{3pt}\noindent\textbf{Code provider}. The code provider (owner) can be the service provider (Scenario 1 in Section~\ref{subsec-scenarios}), and in this case, her target binary (the service code) can be directly handed over to the bootstrap enclave for compliance check. In general, however, the code provider is a different party and may not trust the service provider. So, similar to the data owner, she can also request a remote attestation to verify the bootstrap enclave before delivering her binary to the enclave for a compliance check.


\subsection{Application Scenarios}\label{subsec-scenarios}

The CAT model can be applied to the following scenarios to protect both data and code privacy in computing.  

\begin{figure*}[htbp]
\begin{center}
\centerline{\includegraphics[scale=0.70]{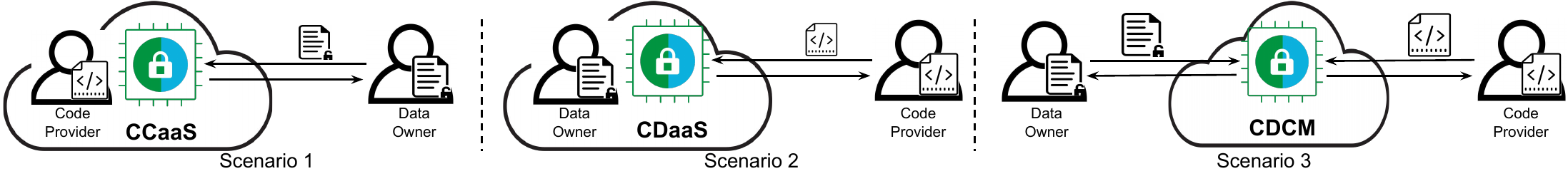}}
\caption{Scenarios}\label{fg-all-scenarios}
\end{center}
\end{figure*}



\vspace{3pt}\noindent\textbf{Scenario~1: Confidential Computing as a Service}.  We consider confidential computing as a service (\textit{CCaaS}) as a privacy extension of today's online data processing services like machine-learning as a service~\cite{russinovich2017introducing}, as the example presented at the beginning of the section. CCaaS is hosted by the party that operates its own\ignore{ or another code provider's} target binary on the data provided by its owner (e.g., an online image classifier to label uploaded user photos). \textit{The outcome of the computation will be sent back to the data owner}. Here, the target binary cannot be released for verification so needs to go through an in-enclave compliance check.



\vspace{3pt}\noindent\textbf{Scenario~2: Confidential Data as a Service}. In this scenario (\textit{CDaaS}), it is the data owner who hosts the online service. The code provider dispatches her program (the target binary) to analyze the data and get the result back, all through a secure channel. An example is that a pharmaceutical company inspects the electronic medical records on a hospital's server to seek suitable candidates for a drug trial. Here, the code provider wants to ensure that her algorithm will be properly executed and will not be released, which is done through a remote attestation to verify the bootstrap loader. The data owner also needs to put a policy in place to control the amount of information that can be given to the code provider.  


\vspace{3pt}\noindent\textbf{Scenario~3: Confidential Data Computing Market}. Another scenario (called \textit{CDCM}) is that the enclave is hosted by an untrusted third party, a market platform, to enable data sharing and analysis. In this case, both the data owner and the code provider upload to the platform their individual content (data or code) through secure channels. They all go through remote attestations to ensure the correctness of the bootstrap enclave, which could also arrange payment transactions between the data owner and the code provider through a smart contract.   




\subsection{Requirements for a CAT System}
\label{subsec-challenges} 

To instantiate the CAT model on a real-world TEE such as SGX, we expect the following requirements to be met by the design: 


\vspace{3pt}\noindent\textbf{Minimizing TCB}.\label{challenge-tcb} 
    In the CAT model the bootstrap enclave is responsible for enforcing security and privacy policies and for controlling the interfaces that import and export code/data for the enclave. So it is critical for trust establishment and needs to be kept as compact as possible for code inspection or verification.


\vspace{3pt}\noindent\textbf{Reducing resource consumption}.\label{challenge-size} 
Today's TEEs operate under resource constraints.  Particularly, SGX is characterized by limited EPC. To maintain reasonable performance, we expect that the software stack of the CAT model controls its resource use. 


\vspace{3pt}\noindent\textbf{Controlling dynamic code loading}.\label{challenge-dep} The target binary is dynamically loaded and inspected by the bootstrap enclave. However, the binary may further sideload other code during its runtime. Some TEE hardware, SGX in particular, does not allow dynamic change to enclave page's RWX properties. So the target binary, itself loaded dynamically, is executed on the enclave's heap space. Preventing it from sideloading requires a data execution prevention (DEP) scheme to guarantee the W $\oplus$ X privilege.

    
   
\vspace{3pt}\noindent\textbf{Preventing malicious control flows}.\label{challenge-cfi} 
    Since the target binary is not trusted, the CAT software stack should be designed to prevent the code from escaping policy enforcement by redirecting its control flow or tampering with the bootstrap enclave's critical data structures. Particularly, previous work shows that special SGX instructions like ENCLU could become unique gadgets for control flow redirecting~\cite{biondo2018guard}, which therefore need proper protection. 

\vspace{3pt}\noindent\textbf{Minimizing performance impact}.\label{challenge-perf} In all application scenarios, the data owner and the code provider expect a quick turnaround from code verification. Also the target binary's performance should not be significantly undermined by the runtime compliance check. 


\subsection{Threat Model}
\label{subsec-threat}

The CAT model is meant to establish trust between the enclave and the code provider, as well as the data owner, under the following assumptions: 

 
\vspace{2pt}\noindent$\bullet$ We do not trust the target binary (service code) and the platform hosting the enclave. In CCaaS, the platform may deliberately run vulnerable target binary to exfiltrate sensitive data, by exploiting the known vulnerabilities during computation. The binary can also leak the data through a covert channel (e.g., page fault~\cite{xu2015controlled}).   


\vspace{2pt}\noindent$\bullet$ Under the untrusted service provider, our model does not guarantee the correctness of the computation, since it is not meant to inspect the functionalities of the target binary. Also, although TEE is designed to prevent information leaks to the untrusted OS, denial of service can still happen, which is outside the scope of the model. 


\vspace{2pt}\noindent$\bullet$ We assume that the code of the bootstrap enclave can be inspected to verify its functionalities and correctness.  Also we consider the TEE hardware, its attestation protocol, and all underlying cryptographic primitives to be trusted.  

\vspace{2pt}\noindent$\bullet$ Our model is meant to protect data and code against different kinds of information leaks, not only explicit but also implicit.  However, side channel for a user-land TEE (like SGX) is known to be hard to eliminate. So our design for instantiating the model on SGX (Section~\ref{subsec-policies}) can only mitigate some types of side-channel threats.  





\ignore{

\section{Confidential Attestation}\label{sec-CAT}

\subsection{Motivating Example}\label{subsec-motivation}

Consider a company providing data-processing services, such as image editing (Pixlr), tax preparation (TurboTax), personal health analysis (23andMe) and deep learning inference as a service, the users need to disclose their sensitive data, such as images, tax documents, and health data, to leverage the convenience and expertise of these services. The user can acquire the data processing performed inside an SGX enclave, and further, verify the enclave code through remote attestation to prevent unintended data disclosure. However, the service provider may not want to disclose the enclave code to the user for verification due to intellectual property reasons, in which case the SGX attestation model fails to fulfill the privacy requirement of both the user (i.e. the data owner) and the service provider.

In this section, we propose the \textit{Confidential ATtestation} (CAT) model to allow the data owner to verify that the enclave code satisfies predefined privacy and security policies without undermining the privacy of the enclave code.
We enumerate several application scenarios that could benefit from the proposed CAT model. We will further list the design goals in designing a practical CAT model in the context of Intel SGX, as well as the threat model considered in the paper. 




\subsection{The CAT model}

There are 4 parties involved in the CAT model.


\vspace{3pt}\noindent\textbf{The IAS}. We assume a standard IAS as in Intel's attestation model who verifies the quote and generates the attestation report. 

\vspace{3pt}\noindent\textbf{The bootstrap enclave}. It is the initial code built into the enclave, which is measured by the hardware to generate the measurement report. It is responsible to authenticate and dynamically load the code and data for the computation to be conducted. It then verifies the code and data is compliant to predefined policies and bootstraps the computation. It is responsible for establishing protected communication channels, and audit the ECalls/OCalls to sanitize the data sent to or received from outside of the enclave.

\vspace{3pt}\noindent\textbf{The data providers}. Depending on the applications, there can be multiple data providers who feed data into the enclave and would like to keep the data from leaking out of the enclave. For this purpose they initiate a remote attestation by sending an attestation request challenge to the enclave developer, inspect and verify the code of the bootstrap enclave by checking the remote attestation report. If they are convinced the bootstrap enclave can guarantee the predefined policies are applied to the loaded code, they can send the data (in encrypted form) to the bootstrap enclave.

\vspace{3pt}\noindent\textbf{The code providers}. Similar as the data providers, code provider/owner could verify the code of the bootstrap enclave through remote attestation. Only after they believe that the bootstrap enclave guarantees the predefined policies, the code can be sent into the bootstrap enclave secretly.

\subsection{Application Scenarios}\label{subsec-scenarios}

In the following, we enumerate several application scenarios that could benefit from the proposed CAT model. 

\begin{figure}[htbp]
\begin{center}
\centerline{\includegraphics[scale=0.72]{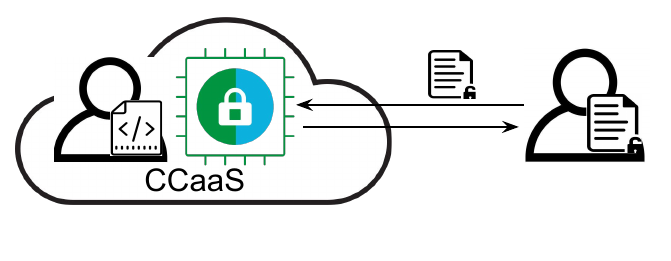}}
\caption{Scenario 1}\label{fg-scenario_1}
\end{center}
\end{figure}

\weijie{change the name of scenarios to: confidential computing as a service, confidential data as a service, etc.}

\vspace{3pt}\noindent\textbf{Scenario~\ref{fg-scenario_1}: Privacy preserving online data processing service}. As the motivating example (Sec.~\ref{subsec-motivation}) shows, the data-processing service (i.e., the bootstrap enclave) is hosted on an SGX-enabled platform owned by the service provider (i.e., the code provider), while the user (i.e., the data provider) sends her encrypted sensitive data for processing. The result will be sent back to the data provider (possibly in encrypted form). As such, the user would like that the bootstrap enclave can enforce the security policy that her data would never leave the enclave.

\begin{figure}[htbp]
\centerline{\includegraphics[scale=0.72]{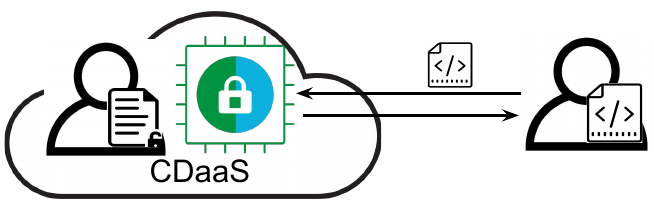}}
\caption{Scenario 2}\label{fg-scenario_2}
\end{figure}

\vspace{3pt}\noindent\textbf{Scenario~2: Privacy preserving online data as a service}. In this scenario, the data-as-a-service (i.e., the bootstrap enclave) is hosted on an SGX-enabled platform owned by the data provider. A user (i.e., the code owner) who would like to conduct computation (such as genome analysis) on the data sends her sensitive code via a secure channel . The result will be sent back to the code provider (possibly in encrypted form). As such, the user would like that the bootstrap enclave can enforce the security policy that the data used are not faked or impure, and her code will not be unintended redirected so that the result is reliable.

\begin{figure}[htbp]
\centerline{\includegraphics[scale=0.72]{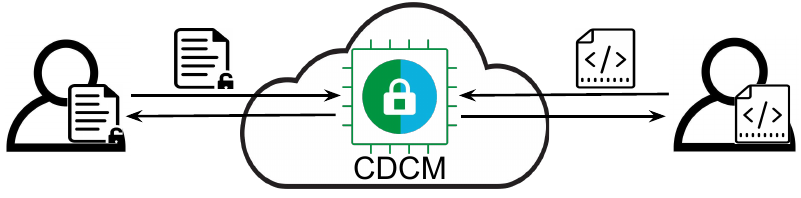}}
\caption{Scenario 3}\label{fg-scenario_3}
\end{figure}

\vspace{3pt}\noindent\textbf{Scenario~3: Privacy preserving online data market}. In this scenario, a data market platform (i.e., the bootstrap enclave) is hosted on a third party.
The data owner uploads her encrypted data to the platform so that she gets paid if the data is used. The data user uploads her code (e.g. genome analysis) also secretly to be computed on people's data satisfying some preset conditions. As such, besides enforcing the confidentiality of both the code and data, the data owner would like to ensure that she gets paid as long as the data is used, while the data user want to ensure that she's not overcharged.

Besides the above scenarios, CAT may be used in more applications, such as privacy preserving smart contract, etc.

\subsection{Design goals}
\label{subsec-challenges} 
At the core of instantiating the CAT model is the design of the bootstrap enclave who is responsible for enforcing the privacy and security policies.
We list the design goals to enable a secure and efficient bootstrap enclave.

\vspace{3pt}\noindent\textbf{Minimizing the Trusted Computing Base (TCB)}.\label{challenge-tcb} 
    In the CAT model the bootstrap enclave is responsible for enforcing the security and privacy policies and to control the interfaces between the loaded code/data and outside of the enclave. The trust built by the CAT model will collapse once it is compromised. It is essential to control the size of the bootstrap enclave and that it can be (formally) verified in the future.

\vspace{3pt}\noindent\textbf{Reducing the memory consumption}.\label{challenge-size} 
    There are limited EPC memory that could be used by the SGX enclave. Considering that the data-processing computation itself consume considerable memory, the design needs to reduce the memory cost adequately.

\vspace{3pt}\noindent\textbf{Confining the (untrusted) loaded code}.\label{challenge-dep} On current SGX hardware, dynamically changing the RWX properties of enclave pages are not supported. So the loaded program will be executed on enclave's heap space, and we need a fine data execution prevention (DEP) scheme to guarantee W $\oplus$ X privilege.
    
   
\vspace{3pt}\noindent\textbf{Preventing malicious control flows}.\label{challenge-cfi} 
    Previous work shows that special instructions in SGX like ENCLU would make unique gadgets for control flow redirecting attacks.
    Since the loaded code can not be trusted, the design needs to prevent code from escaping the policy enforcement by redirecting the control flow and tampering the security-sensitive data structures of the bootstrap code.

\vspace{3pt}\noindent\textbf{Performance considerations}.\label{challenge-perf} It is also important in many scenarios to reduce the time for checking the policy compliance and to induce low run-time overhead for the computation. 

\subsection{Threat model}
\label{subsec-threat}
To demonstrate how to instantiate the CAT model, in the rest of the paper we consider a specific application scenario, i.e., privacy preserving online data processing (Scenario 1 in Sec.~\ref{subsec-scenarios}). As shown in Sec.~\ref{subsec-motivation}, many real world privacy preserving data processing tasks fall into this scenario. 
As described earlier for Scenario 1, the user (i.e., the data provider) submits her sensitive data to a service provider (i.e., the code owner) for data processing tasks. In our threat model, we make the following assumptions.
 

\vspace{2pt}\noindent$\bullet${ The service code and the SGX-enabled host platform are not trusted.} The service provider may (intentionally) write vulnerable service code which causes the leakage of the users' data, e.g., the enclave may be compromised by another user with memory corruption attacks. The service code can even collude with the SGX-enabled platform owned and controlled by the attacker, e.g., through covert channels (such as page faults etc.). 


\vspace{2pt}\noindent$\bullet$ Since the service provider is untrusted, our design does not intend to guarantee the correctness of the results returned to the users. The service users can refuse to pay for the service or turn to other service providers if the results are inferior. This is similar as Denial-of-Service (DoS) attacks which are also out of scope.

\vspace{2pt}\noindent$\bullet$ Besides, we assume the bootstrap enclave code can be inspected (or formally verified) by the users through remote attestation, so that it is trusted to be functional as designed. We trust the SGX hardware and the Intel's attestation protocol as well as the cryptographic algorithms underneath. The mutual authentication between the users and service provider is orthogonal to the design and is omitted in the paper.

}

\section{Enhancing SGX with CAT}\label{sec-design}


In this section we present our design, called \textit{CAT-SGX}, that elevates the SGX platform with the support for the CAT model. This is done using an in-enclave software layer -- the bootstrap enclave running the code consumer and an out-enclave auxiliary -- the code generator. Following we first describe the general idea behind our design and then elaborate the policies it supports, its individual components and potential extension.

\subsection{CAT-SGX: Overview}
\label{subsec-overview}

\noindent\textbf{Idea}. Behind the design of CAT-SGX is the idea of PCC, which enables efficient in-enclave verification of the target binary's policy compliance on the proof generated for the code. A direct application of the existing PCC techniques, however, fails to serve our purpose, as mentioned earlier, due to the huge TCB introduced, the large proof size and the exponential time with regards to the code size for proof generation. To address these issues, we design a lightweight PCC-type 
approach with an untrusted code producer and a trusted code consumer running inside the bootstrap enclave. The producer compiles the source code of the target program (for service providing), generates a list of its indirect jump targets, and instruments it with security annotations for runtime mediation of its control flow and key operations, in compliance with security policies. The list and security annotations constitute a ``proof'', which is verified by the consumer after loading the code into the enclave and before the target binary is activated. 



\begin{figure}[htbp]
\centerline{\includegraphics[scale=0.45]{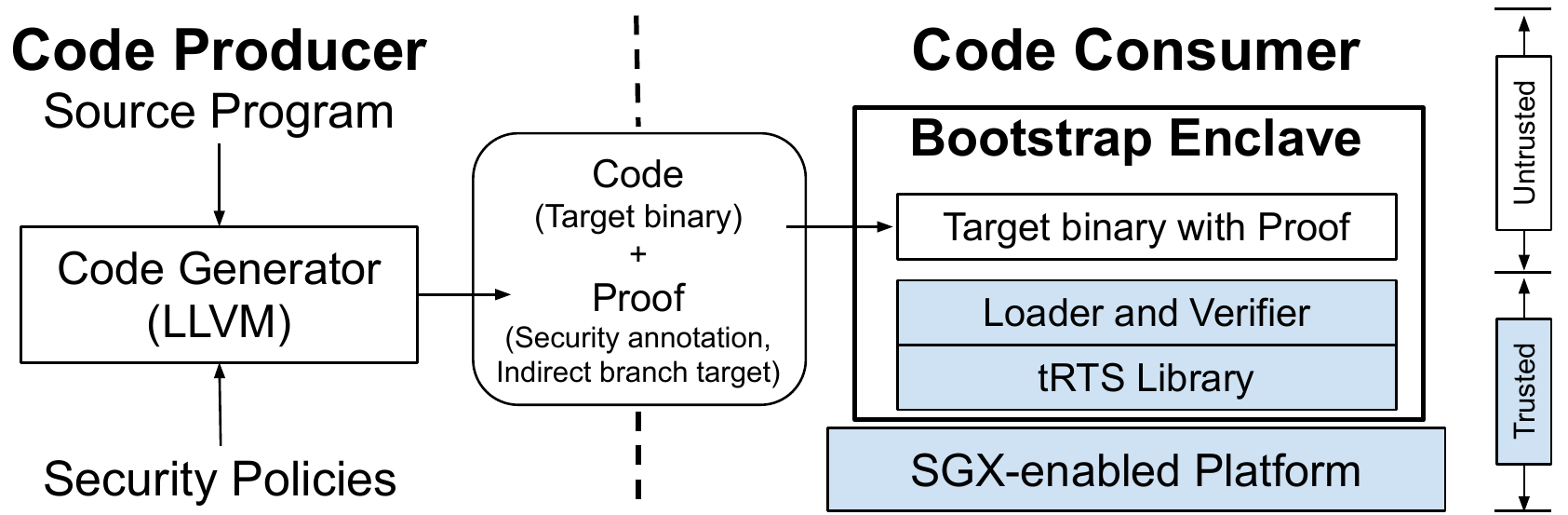}}
\caption{System overview}\label{fg-overview}
\end{figure}

\vspace{3pt}\noindent\textbf{Architecture}. The architecture of CAT-SGX is illustrated in Figure~\ref{fg-overview}. The code generator and the binary and proof it produced are all considered untrusted. Only in the TCB is the code consumer with two components: a dynamic-loader operating a rewriter for re-locating the target binary, and a proof verifier running a disassembler for checking the correct instrumentation of security annotations. These components are all made public and can therefore be measured for a remote attestation (Section~\ref{subsec:ra-impl}). They are designed to minimize their code size, by moving most workload to the code producer. 





\begin{figure*}[htbp]
\centerline{\includegraphics[scale=0.5]{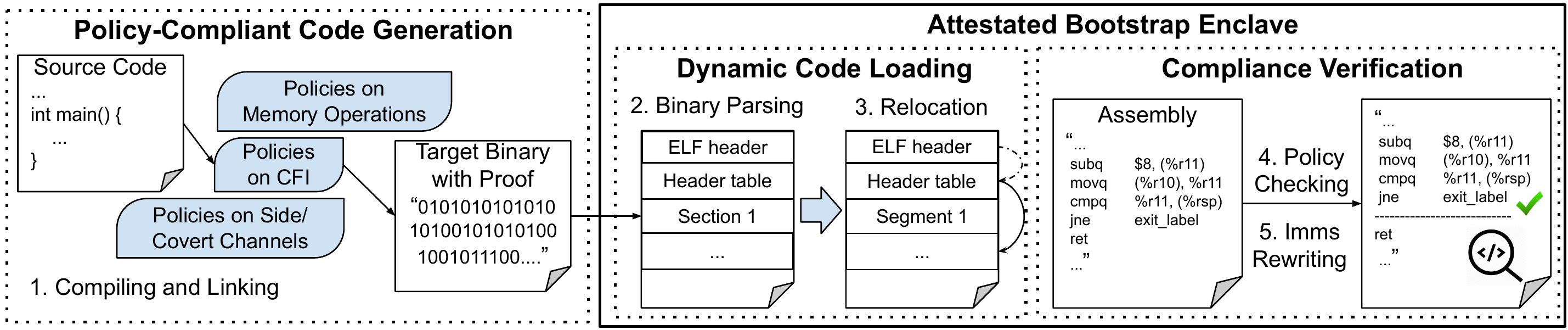}}
\caption{Detailed framework and workflow}\label{fg-workflow}
\end{figure*}

 We present the workflow of CAT-SGX in Figure~\ref{fg-workflow}. 
The target program (the service code) is first instrumented by the code producer, which runs a customized LLVM-based compiler (step 1). 
Then the target binary with the proof (security annotations and the jump target list) are delivered to the enclave through a secure channel.  The code is first parsed (step 2) and then disassembled from the binary's entry along with its control flow traces.
After that, the proof with the assembly inspected by the verifier and if correct (step 3) before some immdiates being rewriten (step 4), is further relocated and activated by the dynamic loader (Step 5). Finally, after the bootstrap transfers the execution to the target program, the service begins and policies are checked at runtime.

\subsection{Security Policies}\label{subsec-policies}

Without exposing its code for verification, the target binary needs to be inspected for compliance with security policies by the bootstrap enclave. These policies are meant to protect the privacy of sensitive data, to prevent its unauthorized disclosure. The current design of CAT-SGX supports the policies in the following five categories: 


\vspace{3pt}\noindent\textbf {Enclave entry and exit control}. CAT-SGX can mediate the content imported to or exported from the enclave, through the ECall and OCall interfaces, for the purposes of reducing the attack surface and controlling information leaks. 


\vspace{2pt}\noindent$\bullet$\textit{ P0: Input constraint, output encryption and entropy control}.
We restrict the ECall interfaces to just serving the purposes of uploading data and code, which perform authentication, decryption and optionally input sanitization (or a simple length check). Also only some types of system calls are allowed through OCalls. Particularly, all network communication through OCalls should be encrypted with proper session keys (with the data owner or the code provider). For CCaaS, the data owner can demand that only one OCall (for sending back results to the owner) be allowed. For CDaaS, the data owner can further impose the constraint on the amount of information (number of bits) that can be returned to the code provider: e.g., one bit to indicate whether suitable patients for a drug trial exist or one byte to tell the number.   


\vspace{3pt}\noindent\textbf{Memory leak control}. Information leak can happen through unauthorized write to the memory outside the enclave, which should be prohibited through the code inspection. 


\vspace{2pt}\noindent$\bullet$\textit{ P1: Preventing explicit out-enclave memory stores}. This policy prevents the target binary from writing outside the enclave, which could be used to expose sensitive data. It can be enforced by security annotations through mediation on the destination addresses of memory store instructions (such as \texttt{MOV}) to ensure that they are within the enclave address range \texttt{ELRANGE}).

\vspace{2pt}\noindent$\bullet$\textit{ P2: Preventing implicit out-enclave memory stores}. Illicit RSP register save/spill operations can also leak sensitive information to the out-enclave memory by pushing a register value to the address specified by the stack pointer, which is prohibited through inspecting the RSP content.  
    
\vspace{2pt}\noindent$\bullet$\textit{ P3: Preventing unauthorized change to security-critical data within the bootstrap enclave}. This policy ensures that the security-critical data would never be tampered with by the untrusted code.



\vspace{2pt}\noindent$\bullet$\textit{ P4: Preventing runtime code modification}. Since the target code is untrusted and loaded into the enclave during its operation, under SGXv1, the code can only be relocated to the pages with \texttt{RWX} properties. So software-based DEP protection should be in place to prevent the target binary from changing itself or uploading other code at runtime. 



\vspace{3pt}\noindent\textbf{Control-flow management}. 
To ensure that security annotations and other protection cannot be circumvented at runtime, the control flow of the target binary should not be manipulated. For this purpose, the following policy should be enforced:  


\vspace{2pt}\noindent$\bullet$\textit{ P5: Preventing manipulation of indirect branches to violate policies P1 to P4}. This policy is to protect the integrity of the target binary's control flow, so security annotations cannot be bypassed. To this end, we need to mediate all indirect control transfer instructions, including indirect calls and jumps, and return instructions.


\vspace{3pt}\noindent\textbf{AEX based side/covert channel mitigation}. SGX's user-land TEE design exposes a large side-channel surface, which cannot be easily eliminated. In the meantime, prior research shows that many side-channel attacks cause Asynchronous Enclave Exits (AEXs). Examples include the controlled side channel attack~\cite{xu2015controlled} that relies on triggering page faults, and the attacks on L1/L2 caches~\cite{wang2017leaky}, which requires context switches to schedule between the attack thread and the enclave thread, when Hyper-threading is turned off or a co-location test is performed before running the binary~\cite{chen2018racing}. CAT-SGX is capable of integrating existing solutions to mitigate the side- or covert-channel attacks in this category.




\vspace{2pt}\noindent$\bullet$\textit{ P6: Controlling the AEX frequency}. The policy requires the total number of the AEX concurrences to keep below a threshold during the whole computation. Once the AEX is found to be too frequent, above the threshold, the execution is terminated to prevent further information leak.

\subsection{Policy-Compliant Code Generation}
\label{subsec-producer}

As mentioned earlier, the design of CAT-SGX is to move the workload from in-enclave verification to out-enclave generation of policy-compliant binary and its proof (security annotations and the list of indirect jump targets). 
In this section we describe the design of the code generator, particularly how it analyzes and instruments the target program so that security policies (P1\textasciitilde P6, see Section~\ref{subsec-policies}) can be enforced during the program's runtime. Customized policies for purposes other than privacy can also be translated into proof and be enforced flexibly.


 


\vspace{3pt}\noindent\textbf{Enforcing P1}.  The code generator is built on top of the LLVM compiler framework (Section~\ref{subsec-instrument}). When compiling the target program (in C) into binary, the code generator identifies (through the LLVM API \verb|MachineInstr::mayStore()|) all memory storing operation instructions (e.g., \texttt{MOV}, Scale-Index-Base (SIB) instructions) and further inserts annotation code before each instruction to check its destination address and ensure that it does not write outside the enclave at runtime. 
The boundaries of the enclave address space can be obtained during dynamic code loading, which is provided by the loader (Section~\ref{subsec:verify}). The correct instrumentation of the annotation is later verified by the code consumer inside the enclave. 


\vspace{3pt}\noindent\textbf{Enforcing P2}. 
The generator locates all instructions that explicitly modify the stack pointer (the RSP in X86 arch) from the binary (e.g., a \texttt{MOV} changing its content) and inserts annotations to check the validity of the stack pointer after them. This protection, including the content of the annotations and their placement, is verified by the code consummer (Section~\ref{subsec-loading}). 
Note that RSP can also be changed implicitly, e.g., through pushing oversized objects onto the stack. This violation is prevented by the loader (Section~\ref{subsec-loading}), which adds guard pages (pages without permission) around the stack. 


\vspace{3pt}\noindent\textbf{Enforcing P3}. Similar to the enforcement of P1 and P2, the code generator inserts security annotations to prevent (both explicit and implicit) memory write operations on security-critical enclave data (e.g., SSA/TLS/TCS) once the untrusted code is loaded and verified. These annotation instructions are verified later by the verifier. 

\vspace{3pt}\noindent\textbf{Enforcing P4}. To prevent the target binary from changing its own code at runtime, the code generator instruments all its write operations (as identified by the APIs \verb|readsWritesVirtualRegister()| and \verb|mayStore()|) with the annotations that disallow alternation of code pages. Note that the code of the target binary has to be placed on \texttt{RWX} pages by the loader under SGXv1 and its stack and heap are assigned to \texttt{RW} pages (see Sec.~\ref{subsec-loading}), so runtime code modification cannot be stopped solely by page-level protection (though code execution from the data region is defeated by the page permissions).


\vspace{3pt}\noindent\textbf{Enforcing P5}. To control indirect calls or indirect jumps in the target program, the code generator extracts all labels from its binary during compilation and instruments security annotations before related instructions to ensure that only these labels can serve as legitimate jump targets. The locations of these labels should not allow an instrumented security annotations to be bypassed. 
Also to prevent the backward-edge control flow manipulation (through \texttt{RET}), the generator injects annotations after entry into and before return from every function call to operate on a shadow stack (see Figure~\ref{fg-shadowstack}), which is allocated during code loading. Also all the legitimate labels are replaced by the loader when relocating the target binary. Such annotations are then inspected by the verifier when disassembling the binary to ensure that protection will not be circumvented by control-flow manipulation (Section~\ref{subsec-disassembling}).

\vspace{3pt}\noindent\textbf{Enforcing P6 with SSA inspection}. When an exception or interrupt take place during enclave execution, an AEX is triggered by the hardware to save the enclave context (such as general registers) to the state saving area (SSA). This makes the occurrence of the AEX visible~\cite{gruss2017strong,chen2018racing}. Specifically, the code generator can enforce the side-channel mitigation policy by instrumenting every basic block with an annotation that sets a marker in the SSA and monitors whether the marker is overwritten, which happens when the enclave context in the area has been changed, indicating that an AEX has occurred. Through counting the number of consecutive AEXes, the protected target binary can be aborted in the presence of anomalously frequent interrupts. This protection can also be verified by the code consumer before the binary is allowed to run inside the enclave.

\vspace{3pt}\noindent\textbf{Code loading support}.\label{subsec:code-loading-support}
Loading the binary is a procedure that links the binary to external libraries and relocates the code. 
For a self-contained function (i.e., one does not use external elements), compiling and sending the bytes of the assembled code is enough. However, if the function wants to use external elements but not supported inside an enclave (e.g., a system call), a distributed code loading support mechanism is needed. In our design, the loading procedure is divided into two parts, one (linking) outside and the other (relocation) inside the enclave.

Our code generator assembles all the symbols of the entire code (including necessary libraries and dependencies) into one relocatable file via static linking. While linking all object files generated by the LLVM, it keeps all symbols and relocation information held in relocatable entries. 
The relocatable file, as above-mentioned target binary, is expected to be loaded for being relocated later (Section~\ref{subsec-loading}).







\subsection{Configuration, Loading and Verification}
\label{subsec:verify}

With the annotations instrumented and legitimate jump targets identified, the in-enclave workload undertaken by the bootstrap enclave side has been significantly reduced. Still, it needs to be properly configured to enforce the policy (P0) that cannot be implemented by the code generator, load and relocate the target binary so instrumented protection can be properly executed and also verify the ``proof'' for policy compliance through efficient dissembling and inspecting the binary. Following we elaborate how these critical operations are supported by our design.


\vspace{3pt}\noindent\textbf{Enclave configuration to enforce P0}. 
To enforce the input constraint, we need to configure the enclave by defining certain public ECalls in Enclave Definition Language (EDL) files for data and code secure delivery. Note such a configuration, together with other security settings, can be attested to the remote data owner or code provider. The computation result of the in-enclave service is encrypted using a session key (with the data owner or code provider) after the remote attestation and is sent out through a customized OCall. For this purpose, CAT-SGX only defines allowed system calls (e.g., \texttt{send/recv}) in the EDL file, together with their wrappers for security control. Specially, the wrapper for \texttt{send} encrypts the message to be delivered and pads it to a fixed length. 

To support the CCaaS setting, only \texttt{send} and \texttt{recv} are allowed to communicate with the data owner. When necessary, the wrappers of these functions can pad the encrypted output and ensure that the inter-packet timings are constant to mitigate the side-channel risk. For CDaaS, we only permit a \texttt{send} OCall to be invoked once to deliver the computing result to the code provider, which can be enforced by the wrapper of the function through a counter. Further the wrapper can put a constraint on the length of the result to control the amount of information disclosed to the code provider: e.g., only 8 bits can be sent out. 

\vspace{3pt}\noindent\textbf{Dynamic code loading and unloading}. \label{subsec-loading}
The target binary is delivered into the enclave as data through an ECall, processed by the wrapper placed by CAT-SGX, which authenticates the sender and then decrypts the code before handing it over to the dynamic loader. The primary task of the loader is to rebase all symbols of the binary according to its relocation information (Section~\ref{subsec:code-loading-support}). For this purpose, the loader first parses the binary to retrieve its relocation tables,  then updates symbol offsets\ignore{ based upon the symbol tables}, and further reloads the symbols to designated addresses. During this loading procedure, the indirect branch label list is ``translated'' to in-enclave addresses, which are considered to be legitimate branch targets and later used for policy compliance verification. 


As mentioned earlier (Section~\ref{subsec-producer}), the code section of the target binary is placed on pages with \texttt{RWX} privileges, since under SGXv1, the page permissions cannot be changed during an enclave's operation, while the data sessions (stack, heap) are assigned to the pages with \texttt{RW} privileges. These code pages for the binary are guarded against any write operation by the annotations for enforcing P4. Other enclave code, including that of the code consumer, is under the \texttt{RX} protection through enclave configuration. Further the loader assigns two non-writable blank guard pages right before and after the target binary's stack for enforcing P2, and also reserves pages for hosting the list of legitimate branch targets and the shadow stack for enforcing P5.


\vspace{3pt}\noindent\textbf{Just-enough disassembling and verification}.
\label{subsec-disassembling} After loading and relocating, the target binary is passed to the verifier for a policy compliance check. Such a verification is meant to be highly efficient, together with a lightweight disassembler. Specifically, our disassembler is designed to leverage the assistance provided by the code generator.  It starts from the program entry discovered by the parser and follows its control flow until an indirect control flow transfer, such as indirect jump or call, is encountered. Then, it utilizes all the legitimate target addresses on the list to continue the disassembly and control-flow inspection. In this way, the whole program will be quickly and comprehensively examined.  

For each indirect branch, the verifier checks the annotation code (Figure~\ref{subsec-producer}) right before the branch operation, which ensures that the target is always on the list at runtime. Also, these target addresses, together with direct branch targets, are compared with all guarded operations in the code to detect any attempt to evade security annotations. With such verification, we will have the confidence that no hidden control transfers will be performed by the binary, allowing further inspection of other instrumented annotations. These annotations are expected to be well formatted and located around the critical operations as described in Section~\ref{subsec-producer}. Figure~\ref{fg-mov} presents an example and more details are given in Section~\ref{subsec-instrument} and Appendix.

\ignore{

\vspace{3pt}\noindent\textbf{Enforcing P0}. 
To enforce the input constraint, we define certain public ECalls in Enclave Definition Language (EDL) files for data and code secret delivery, which can be attested by the remote data owner.
The computation result of the in-enclave service is encrypted using session key after RA and will be output in a customized OCall. Before that, to construct the output, the messages are padded to a constant length.  
In this CCaaS setting, only one Ocall (for outputing the computation result) is allowed to execute once. In scenarios like serving an HTTPS server and other network services, only the send/receive Ocall interfaces are allowed, and the interval time of each send/receive can be padded to prevent the timing covert channels.

\vspace{3pt}\noindent\textbf{Enforcing P1}. The code generator finds all memory storing operations and instruments annotation code before them to make sure they are writing to the address space out of the enclave. 
The boundaries of the enclave address space can be obtained during the procedure of dynamic code loading (Sec.~\ref{subsec:verify}).
The code consumer the enclave confirms such security check instructions are in place.

\vspace{3pt}\noindent\textbf{Enforcing P2}. 
The generator locates all the instructions modifying the stack pointer (aka. the RSP in X86 arch) explicitly, and inserts instructions to check the validity of the stack pointer after them. The code consumer confirms the placement of these check instruction. Furthermore, the code consumer prevents implicit modification (pushing oversized objects to the stack) to the stack pointer by adding guard pages (i.e., pages granted no permissions) around the stack boundaries. 

\vspace{3pt}\noindent\textbf{Enforcing P3}. Similar to enforcing P1 and P2, the code generator further enforces that (both explicit and implicit) memory write operations cannot alter the security-critical data once the untrusted code is loaded and verified. These annotation instructions are verified later in the verifier. 

\vspace{3pt}\noindent\textbf{Enforcing P4}. Similar to enforcing P1 and P2, the code generator and consumer enforce that memory write operations cannot modify the \texttt{RWX} pages. We can combine the enforcement of from P1 to P4, setting several boundaries for all of them.

\vspace{3pt}\noindent\textbf{Enforcing P5}. For indirect calls or indirect jumps, the code generator firstly extract all legal destination addresses, then store them as a list in a specific reserved region. It instruments code to check the targets of all indirect call/jump instructions in the code to ensure they only direct to addresses on that list. A shadow stack is also included, to prevent backward-edge control flow manipulation.
Instruments can be efficiently verified, while the integrity of the forward-edge indirect branches will be checked during disassembling (Section~\ref{subsec-disassembling}).


\vspace{3pt}\noindent\textbf{Enforcing P6 by detecting page faults with TSX support (P6-TSX)}.
We enforce P6 by introducing the idea of T-SGX~\cite{shih2017t}. As a compiler-level scheme that automatically transforms a normal enclave program into a secured one, T-SGX can isolate the fallback handler and other transaction control code, called springboard, from the original program’s code and data pages to ensure that exceptions including page faults and timer interrupts can only be triggered on the springboard. We take advantage of it and implement instruction wrappers that encompass all boundaries between any basic blocks and branches. The fallback route of the TSX wrapper records the number of transaction aborts, which ensures that if a threshold is exceeded, the program is forced to exit. 

\vspace{3pt}\noindent\textbf{Enforcing P6 by detecting AEX with monitoring the SSA (P6-SSA)}. When an exception or interrupt is triggered during the enclave execution, the AEX performed by the hardware saves the enclave context (such as general registers) to the state saving area (SSA).
As demonstrated in previous works~\cite{gruss2017strong,chen2018racing}, AEX can be detected by monitoring the SSA. We instrument every basic block to set a marker in the SSA and monitor whether the marker is overwritten by AEX within the basic block. The execution is terminated once the number of AEXs within the basic block exceeds a preset threshold.


\subsection{Code Loading and Compliance Verification}
\label{subsec:verify}

Due to the thorough generated proof with compliance, the bootstrap enclave can quickly verify the validity of the proof, and it can compare the conclusions of the proof to its own security policy to determine whether the application is safe to execute.

\vspace{3pt}\noindent\textbf{Dynamic code loading and unloading}. \label{subsec-dynamicloader}
In our design, the linking procedures (linking and rebasing) of a target program are separated into both inside and outside the enclave respectively. SGX only accepts that code running inside an enclave is linked against the SGX SDK at build time. For a self-contained function (i.e., one does not use external elements), compiling and sending the bytes of the assembled code is enough. However, if the function uses external elements, a distributed mechanism is needed to map these elements into their corresponding positions at the enclave side. So we use separated linking and rebasing to assemble all the symbols of the entire code (including necessary libraries and dependencies) into one relocatable file (aka. linking), and then parse the symbols and load/relocate the code at runtime inside the enclave (aka. rebasing). Further, during the linkage procedure, we also load and relocate an indirect branch target entry list as part of the proof for later runtime verification.

The rebasing process starts with the bootstrap enclave receiving the generated target binary code through a buffer. The dynamic loader's primary task is to rebase all of its symbols according to the information in its relocation table. Therefore, the loader reads relocation tables of the binary, updates symbol offsets of its symbol tables, and reloads symbols to designated addresses. 

\vspace{3pt}\noindent\textbf{Just-enough disassembling and verification}.
\label{subsec-disassembling}
Accurate and complete binary disassembly is a difficult problem in general due to indirect control flow transfers. We built a lightweight disassembler with the assistance of the compiler outside the enclave. Our disassembler starts with the program entry and follows the program control flow. When we encounter indirect control flow transfers such as indirect jumps and indirect calls, we use the valid target list provided by the compiler to find the targets of indirect control flows. Note that our verifier will ensure that there are runtime checks before every indirect control flow, which guarantees that the actual control flow targets are inside the list provided by the compiler. 
When jumping to next target, we check if the target is inside any instrumentations enforced by policies (in Section~\ref{subsec-producer}).
In this way, the untrusted compiler is not able to hide dangerous transfer targets by omitting them in the list. The runtime check ensures the integrity of the target list provided by the compiler.

}

\ignore{

As described in Sec.~\ref{sec-CAT}, a confidential attestation process encompasses a standard Intel attestation to attest the bootstrap enclave and establish a secure communication channel. If they are convinced that the bootstrap enclave can enforce the desired security policies, the data or code providers can send the data and code (in encrypted form) to the bootstrap enclave. 

In this section we present our design for realizing the CAT model under the privacy preserving data processing scenario. In this setting, the service user (i.e., the data owner) communicates with the remote data processing service with encrypted messages for uploading her data and receiving the results.
To minimize the size of the bootstrap enclave, the design borrows the idea of PCC and consists of an untrusted code producer outside the enclave, and a trusted code consumer inside.
Sec.~\ref{subsec-threat} presents the threat model. Sec.~\ref{subsec-policies} presents the categorized security policies enforced by the bootstrap enclave. Sec.~\ref{subsec-producer} and Sec.~\ref{subsec:verify} presents how the policies can be enforced and verified with the cooperative design of the code producer and code consumer.
We will discuss how to extend the design to other scenarios with carefully crafted security policies in Sec.~\ref{subsec:morescenario}.



\ignore{
\subsection{Threat model}
\label{subsec-threat}
To demonstrate how to instantiate the CAT model, in the rest of the paper we consider a specific application scenario, i.e., privacy preserving online data processing (Scenario 1 in Sec.~\ref{subsec-scenarios}). As shown in Sec.~\ref{subsec-motivation}, many real world privacy preserving data processing tasks fall into this scenario. 
As described earlier for Scenario 1, the user (i.e., the data provider) submits her sensitive data to a service provider (i.e., the code owner) for data processing tasks. In our threat model, we make the following assumptions.
 

\vspace{2pt}\noindent$\bullet${ The service code and the SGX-enabled host platform are not trusted.} The service provider may (intentionally) write vulnerable service code which causes the leakage of the users' data.
The service code can even collude with the SGX-enabled platform owned and controlled by the attacker, e.g., through covert channels (such as page faults etc.). 


\vspace{2pt}\noindent$\bullet$ Since the service provider is untrusted, our design does not intend to guarantee the correctness of the results returned to the users. The service users can refuse to pay for the service or turn to other service providers if the results are inferior. This is similar as Denial-of-Service (DoS) attacks which are also out of scope.

\vspace{2pt}\noindent$\bullet$ Besides, we assume the bootstrap enclave code can be inspected (or formally verified) by the users through remote attestation, so that it is trusted to be functional as designed. We trust the SGX hardware and the Intel's attestation protocol as well as the cryptographic algorithms underneath. The mutual authentication between the users and service provider is orthogonal to the design and is omitted in the paper.


}


\subsection{CAT-SGX: Overview}
\label{subsec-overview}

A common method to enable efficient verification of the code properties within the bootstrap enclave is to use PCC. We find, however, applying traditional PCC schemes directly to the CAT model is impractical due to the following reasons: (1) Traditional PCC schemes induce huge TCB, including the VC generator with an intermediate  language interpreter/compiler, or a virtual machine that can support runtime proof checking. (2) In traditional PCC schemes, the proof size is usually large. (3) Although the procedure of generating proof runs outside the enclave and is not trusted, it usually takes exponential time with respect to the code size and may take too long for real world data processing computations.

\begin{figure}[htbp]
\centerline{\includegraphics[scale=0.45]{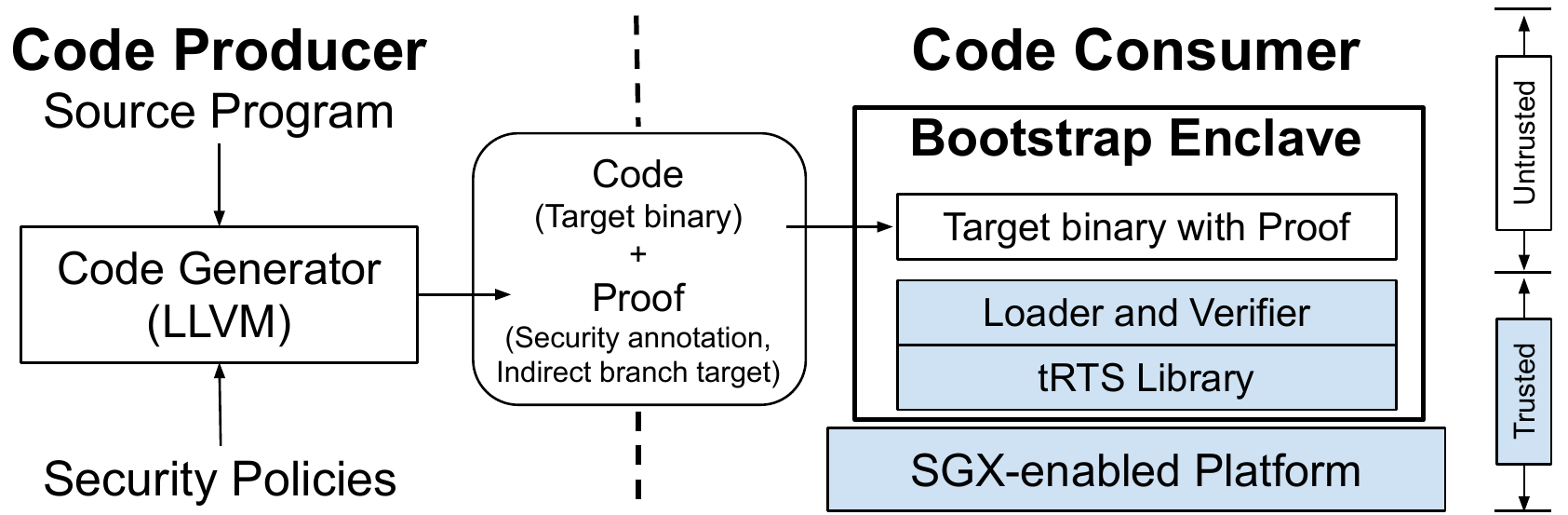}}
\caption{System overview}\label{fg-overview}
\end{figure}

Instead, we construct a lightweight PCC-type scheme which includes an untrusted code producer and a trusted code consumer running in the bootstrap enclave (Fig.~\ref{fg-overview}). The overview design is derived from the original PCC idea while we have simplified it for our own purpose. The illustration of our design lists all components that are divided into two parts, trusted part and untrusted part. In our new PCC-based system, the proof is generated from the outside of the enclave during the compiling and can be verified at runtime inside. The inside verifier can cooperate with the outside compiler to make the verifier as lightweight as possible, using static verification of dynamic checks.

In traditional PCC framework, the VCGen often exists as a compiler~\cite{colby2000certifying,leroy2006formal} or a sandbox~\cite{pirzadeh2010extended}, which is too heavy for limited executive resources. So here, we build our own lightweight PCC system to verify if a cloud service would leakage user’s data. Generally, we provide a code transformer for service code which needs to be verified, and a secure enclave for executing the verified service code. The only TCB in our design is the verifier code inside the enclave, including the dynamic-loader, the disassembler/rewriter (for rewritting structured guards), and the proof verifier (for checking verification hints). On the other hand, the whole trusted code consumer can be remotely attested (in Subsection~\ref{subsec-dynamicloader}), which can indirectly protect the integrity of for code provider, as well as the isolate valuable implementation detail from be accessed.

As mentioned above, to facilitate PCC framework working well in SGX and to make PCC more efficient, specifically to reduce the size of code consumer side, we move as much proof generation workload as possible to the code producer side, and leave as few verification workload as possible to the inside of the enclave.

\begin{figure*}[htbp]
\centerline{\includegraphics[scale=0.5]{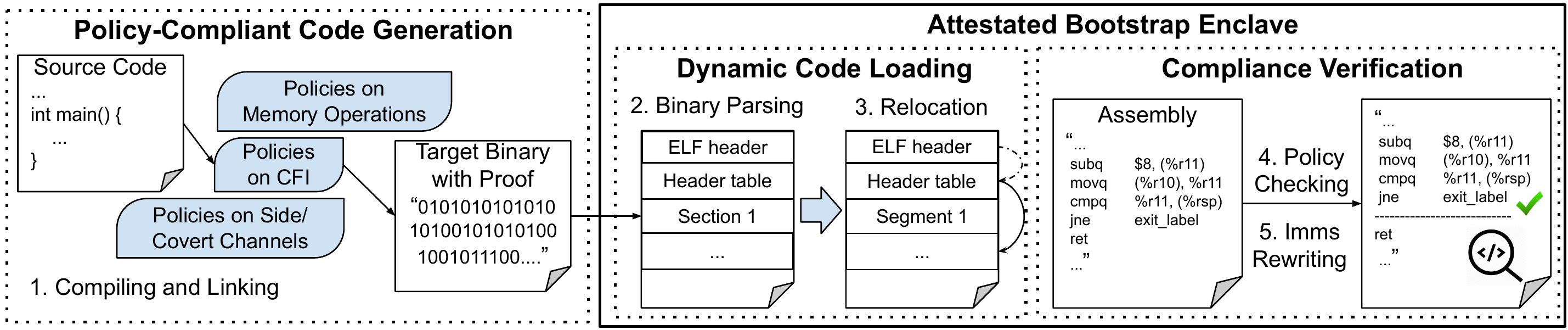}}
\caption{Detailed framework and workflow}\label{fg-workflow}
\end{figure*}

\vspace{3pt}\noindent\textbf{Workflow.} The workflow of our privacy-preserving TEE and PCC framework is shown as Figure~\ref{fg-workflow}. Unlike the typical interaction between the producer and consumer, the workflow encompasses several steps. 

First, the service code (target program) is instrumented using our own `code + proof' generator - a customized LLVM-based compiler (step 1,2). Then on the code consumer side, the loader source code and the verifier source code are compiled with SGX SDK to build the bootstrap enclave binary (step 3). Our loader can load and unload code after initialization (step 4), followed by being remote attested (step 5). Moreover, the `code + proof' can be transferred to code consumer where SGX is deployed, waiting for being dynamic loaded and rebased into the enclave (step 6). After being loaded, the verifier will rewrite some key immediate operands (Imm) and finally transfer the execution to the target program (step 7). 

\vspace{3pt}\noindent\textbf{Dynamic Code Loading and Unloading.} \label{subsec-dynamicloader}
In our design, the linking procedures (linking and rebasing) of a target program are separated into both inside and outside the enclave respectively. SGX only accepts that code running inside an enclave is linked against the SGX SDK at build time. For a self-contained function (i.e., one does not use external elements), compiling and sending the bytes of the assembled code is enough. However, if the function uses external elements, a distributed mechanism is needed to map these elements into their corresponding positions at the enclave side. So we use separated linking and rebasing to assemble all the symbols of the entire code (including necessary libraries and dependencies) into one relocatable file (aka. linking), and then parse the symbols and load/relocate the code at runtime inside the enclave (aka. rebasing). Further, during the linkage procedure, we also load and relocate an indirect branch target entry list as part of the proof for later runtime verification.

The rebasing process starts with the bootstrap enclave receiving the generated binary code through a buffer. The dynamic loader's primary task is to rebase all of its symbols according to information in its relocation table. Therefore, the loader reads relocation tables in the code, updates symbol offsets in symbol tables, and loads symbols to addresses designated in relocation tables. After rebasing, the detailed memory layout is shown on the right side in Fig.~\ref{fg-dynloader}.


\subsection{Security Policies}\label{subsec-policies}

In this scenario, the bootstrap enclave needs to enforce that the data will not be leaked by the untrusted service code, which is not exposed to the data provider. It can be achieved by enforcing several policies as follows. Notably, the policies put some constraints on the service code, yet we make a lot of effort on these constraints to ensure that the code functionality are intact and necessary for the CAT model.

\vspace{3pt}\noindent\textbf {Constraining Ecalls/Ocalls}. In such service-oriented scenarios, all the bridge functions should be public and attestable for normal use, we should make restrictions on them. The output is to be produced encrypted and the loader must deal with system calls via a trusted Ocall routine. All ECalls/OCalls will be audited and configured correctly by the bootstrap enclave.  

\vspace{2pt}\noindent$\bullet$\textit{ P0: Standard and Encrypted Output via legitimate Ocalls}. After the Remote Attestation and the session key exchange, messages sent from the enclave should be all encrypted. And for security consideration, they could be the same length to prevent further inference attacks.

\vspace{3pt}\noindent\textbf {Verifying Memory Operations}. In order to prevent data leakage during it being processed by a untrusted code provider, we need a verifier to check if this program is prone to write sensitive information from inside enclave to the outside world. 

\vspace{2pt}\noindent$\bullet$\textit{ P1: Preventing explicit memory stores to the outside of the enclave}. An enclave has the ability to write data outside of its EPC memory region arbitrarily. Therefore, the major policy is to prevent the untrusted code from copying the data across enclave boundaries. The policy-compliance verifier needs to ensure that the destinations of memory store instructions such as \texttt{MOV} are within the enclave address range (also known as the \texttt{ELRANGE}). 

\vspace{2pt}\noindent$\bullet$\textit{ P2: Preventing implicit memory stores to the outside of the enclave}. Illicit RSP register save/spill operations can do the trick of leaking sensitive information to memory via pushing a register value to an address specified by the stack pointer. 
    
\vspace{2pt}\noindent$\bullet$\textit{ P3: Preventing tampering of security-critical data within the bootstrap enclave}. This ensures that the code never reads or writes secrets in the SSA/TCS area, which is necessary because strong security properties no longer hold if the thread control structure data is used. In this case the verifier needs to ensure that the untrusted code does not tamper with this kind of data structure.

\vspace{3pt}\noindent\textbf{Constraining Control Transfers}. 
In our PCC-type scheme, the policy-compliance verifier performs static analysis of the untrusted code. It is essential to prevent attackers from dynamically redirecting the control flow at runtime, which may bypass the check performed when the code is initially loaded. In this respect, the following policies need to be enforced. 

\vspace{2pt}\noindent$\bullet$\textit{ P4: Data execution prevention for the RWX region}. In the currently SGX platforms that support only SGXv1 instructions, the untrusted code are loaded to pages with \texttt{RWX} properties. A software DEP scheme is needed to prevent the untrusted code from changing the code at runtime.

\vspace{2pt}\noindent$\bullet$\textit{ P5: Indirect branches shall not point to destinations that violate policies P1 to P4}. Such control flow integrity definitely should be guaranteed since the loaded code could be malicious so that it could bypass the mentioned policies. Therefore, this enforcement should be performed for all indirect control transfer instructions, including indirect calls, indirect jumps, and return instructions.

\vspace{3pt}\noindent\textbf{Detecting leakage through AEX based side channels}. Side channels are difficult to eliminate and are severe threats to TEEs such as SGX. As shown in previous works, the abnormal AEXs can be used to detect many low-noise side channels within SGX, such as controlled channel attacks~\cite{xu2015controlled}, and same core L1/L2 cache attacks~\cite{chen2018racing}. We are not meant to design new side channel defenses, nevertheless we propose to transplant 
existing side channel detecting techniques, 
which illustrates the generality of the CAT model. 


\vspace{2pt}\noindent$\bullet$\textit{ Alternative P6: Detecting page faults with TSX support}. 

\vspace{2pt}\noindent$\bullet$\textit{ Alternative P7: Detecting AEX by monitoring the SSA}.

\vspace{3pt}\noindent\textbf{Multi-Usesr Isolation}. When protecting in-enclave time-sharing services, a big challenge is to prevent the service program infected by one user from victimizing another user, and the confidentiality of a user’s data left in the enclave to protect it from leaking to the subsequent user receiving the service.

\vspace{2pt}\noindent$\bullet$\textit{ P8: User data cleansing}. The purpose of the data cleansing and the exit sanitization is to reset the service state and clean up the old user’s data right after execution is done. Except the persistent data of the user, all other data will be cleaned up, together with the content of SSA and registers.

\subsection{Policy-Compliant Code Generation}
\label{subsec-producer}
In this section we present the design of the code generator, which given the source program can produce binary code that enforces the policies P1 to P5, and can be easily verified. The design of the policy verifier will be presented in Sec.~\ref{subsec:verify}. Since we do the verification at the assembly level and the target binary loaded in the bootstrap enclave will finally be disassembled, the code generator does not need to be trusted.

\vspace{3pt}\noindent\textbf{Enforcing P0}. The output message for the in-enclave service is encrypted using session key after RA. Before that, to construct the output message, the plaintext is padded to a constant length if have to. Meanwhile, the output functions are wrapped by our customized Ocall stubs. On the other hand, 37 common system calls are also wrapped with Ocall stubs for possible system interactions.\wenhao{we need to discuss: admitting arbitrary ocalls could lead to information leakage by ocall patterns}

\vspace{3pt}\noindent\textbf{Enforcing P1}. The generator instruments all storing instructions to check if they are writing to the memory out of the enclave. The code consumer inside the enclave confirms such security check instructions are in place.

\vspace{3pt}\noindent\textbf{Enforcing P2}. The generator prevents implicit memory write crossing the enclave boundaries by making sure the stack pointer never point to memory regions outside the enclave. It locates all the instructions modifying the stack pointer explicitly, and inserts instructions to check the validity of the stack pointer after them. The code consumer confirms the placement of these check instruction. Furthermore, the code consumer prevents implicit modification of the stack pointer by adding guard pages (i.e., pages granted no permissions) around the stack boundaries. 

\vspace{3pt}\noindent\textbf{Enforcing P3}. Similar to enforcing P1 and P2, the code generator further enforces that (both explicit and implicit) memory write operations cannot alter the security-critical data once the untrusted code is loaded and verified. These check instructions are verified by the code consumer inside the enclave. 

\vspace{3pt}\noindent\textbf{Enforcing P4}. Similar to enforcing P1 and P2, the code generator and consumer enforce that (both explicit and implicit) memory write operations cannot modify the \texttt{RWX} pages.

\vspace{3pt}\noindent\textbf{Enforcing P5}. For indirect calls or indirect jumps, the code generator firstly extract all legal destination addresses of them. Theses addresses are stored in a specific data region. Then it instruments code to check the targets of all the indirect call/jump instructions in the code to ensure they only direct to addresses listed in the data region. These instruments can be efficiently verified by the in-enclave code consumer.

\vspace{3pt}\noindent\textbf{Enforcing P6}.
We enforce P6 by introducing the idea of T-SGX~\cite{shih2017t}. As a compiler-level scheme that automatically transforms a normal enclave program into a secured one, T-SGX can isolate the fallback handler and other transaction control code, called springboard, from the original program’s code and data pages to ensure that exceptions including page faults and timer interrupts can only be triggered on the springboard. We take advantage of it and implement instruction wrappers that encompass all boundaries between any basic blocks and branches. The fallback route of the TSX wrapper records the number of transaction aborts, which ensures that if a threshold is exceeded, the program is forced to exit. 

\vspace{3pt}\noindent\textbf{Enforcing P7}.
We enforce P7 by retrofitting Hyperrace~\cite{chen2018racing}. However, unlike Hyperrace doing the physical-core co-location tests, here we only monitor how often the interrupts/AEXs happen. L1/L2 cache-based channel can be detected when certain number or more interrupts/AEXs occur in one basic block or every k instructions.

\vspace{3pt}\noindent\textbf{Enforcing P8}. Before the bootstrap enclave is destroyed, all user data is cleared after the execution transferred back to the loader.

\subsection{Compliance Verification}
\label{subsec:verify}

Due to the thorough generated proof with compliance, the bootstrap enclave can quickly verify the validity of the proof, and it can compare the conclusions of the proof to its own security policy to determine whether the application is safe to execute.

\vspace{3pt}\noindent\textbf{Just-enough Disassembling.} Accurate and complete binary disassembly is a difficult problem in general due to indirect control flow transfers. We built a lightweight disassembler with the assistance of the compiler outside the enclave. Our disassembler starts with the program entry and follows the program control flow. When we encounter indirect control flow transfers such as indirect jumps and indirect calls, we use the valid target list provided by the compiler to find the targets of indirect control flows. Note that our verifier will ensure that there are runtime checks before every indirect control flow, which guarantees that the actual control flow targets are inside the list provided by the compiler. In this way, the untrusted compiler is not able to hide dangerous transfer targets by omitting them in the list. The runtime check ensures the integrity of the target list provided by the compiler.
}
\section{Implementation}\label{sec-implementation}

We implemented the prototype on Linux/X86 arch. Specifically, we implemented the code generator with LLVM 9.0.0, and built other parts on an SGX environment.

We implemented one LLVM back-end pass consisting of several types of instrumentations for the code generator, about 1200 lines of C++ code in total. 
Besides, we implemented the bootstrap enclave with over 1900 lines of code based on Capstone~\cite{capstone} as the disassembler. 



\subsection{Assembly-level Instrumentation}\label{subsec-instrument}


\begin{figure}[htbp]
\centerline{\includegraphics[scale=0.38]{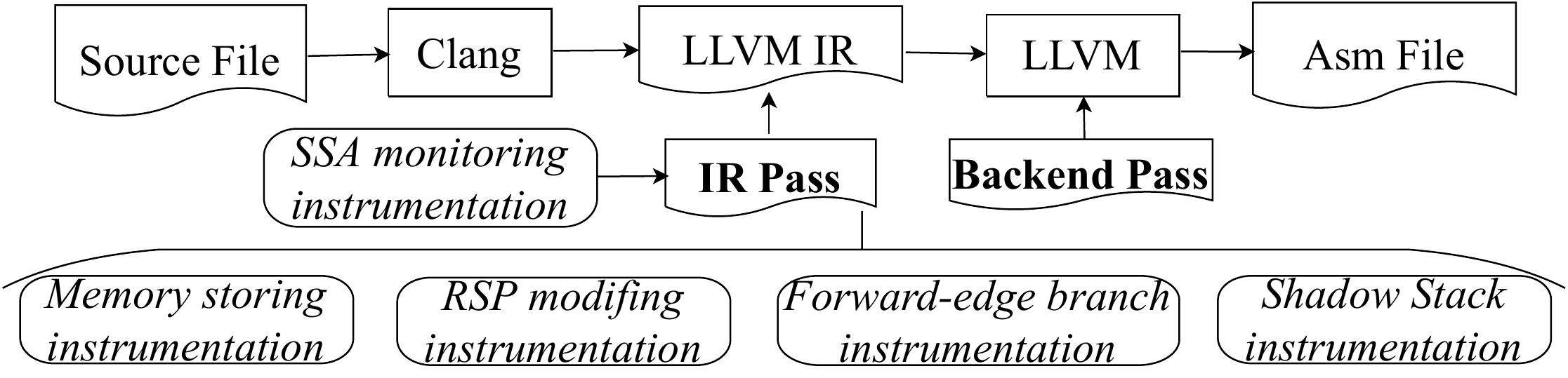}}
\caption{Detailed workflow of the code generator}\label{fg-codegen}
\end{figure}

The code generator we built is mainly based on LLVM (Fig.~\ref{fg-codegen}), and the assembly-level instrumentation is the core module. To address the challenge of limited computing resources described in Section~\ref{challenge-tcb}, this code generator tool is designed and implemented comprehensively, to make the policy verifier small and exquisite. More specifically, we implemented modules for checking memory writing instructions, RSP modification, indirect branches and for building shadow stack. And we reformed a instrumentation module to generate side-channel-resilient annotations. 
Note that we can not only demonstrate the security policies for several real-world scenarios can be efficiently enforced with our framework, modules of the annotation generation for customized functionalities can also be integrated into the code generator.
For convenience, switches to turn on/off these modules are made. 

Here is an example. The main function of the module for checking explicit memory write instructions (P1) is to insert annotations before them. Suppose there is such a memory write instruction in the target program, `\texttt{mov reg, [reg+imm]}',
the structured annotation first sets the upper and lower bounds as two temporary Imms (3ffffffffffff and 4ffffffffffff), and then compares the address of the destination operand with the bounds. The real upper/lower bounds of the memory write instruction are specified by the loader later. If our instrumentation finds the memory write instruction trying to write data to illegal space, it will cause the program to exit at runtime. The code snippet (structured format of the annotation) is shown in Figure~\ref{fg-mov}. More details can be found at Appendix~\ref{appendix-instrumentation}.

\begin{figure}
\begin{center}
\begin{minipage}{0.35\textwidth}
\begin{lstlisting}[basicstyle=\scriptsize]
pushq   %rbx    ;save execution status
pushq   %rax
leaq    [reg+imm], %rax ;load the operand
movq    $0x3FFFFFFFFFFFFFFF, %rbx  ;set bounds
cmpq    %rbx, %rax
ja      exit_label
movq    $0x4FFFFFFFFFFFFFFF, %rbx  ;set bounds
cmpq    %rbx, %rax              
jb      exit_label
popq    %rax
popq    %rbx
\end{lstlisting}
\end{minipage}
\end{center}
\caption{Store instruction instrumentation}\label{fg-mov}
\end{figure}

Although using the code generator we could automatically produce an instrumented object file, we still need to deal with some issues manually that may affect practical usage. As the workflow described in Figure~\ref{fg-workflow}, the first job to make use of CAT system is preparing the target binary.  Service-specific libraries and some dependencies also should be built and linked against the target program (detailed in Appendix~\ref{appendix-preparing}).

\subsection{Building Bootstrap Enclave}\label{subsec:bootstrap-impl}
Following the design in Section~\ref{subsec:verify}, we implemented a \textit{Dynamic Loading after RA mechanism} for the bootstrap enclave. During the whole service, the data owner can only see the attestation messages which are related with the bootstrap's enclave quote, but nothing about service provider’s code.

\begin{figure}[htbp]
\centerline{\includegraphics[scale=0.45]{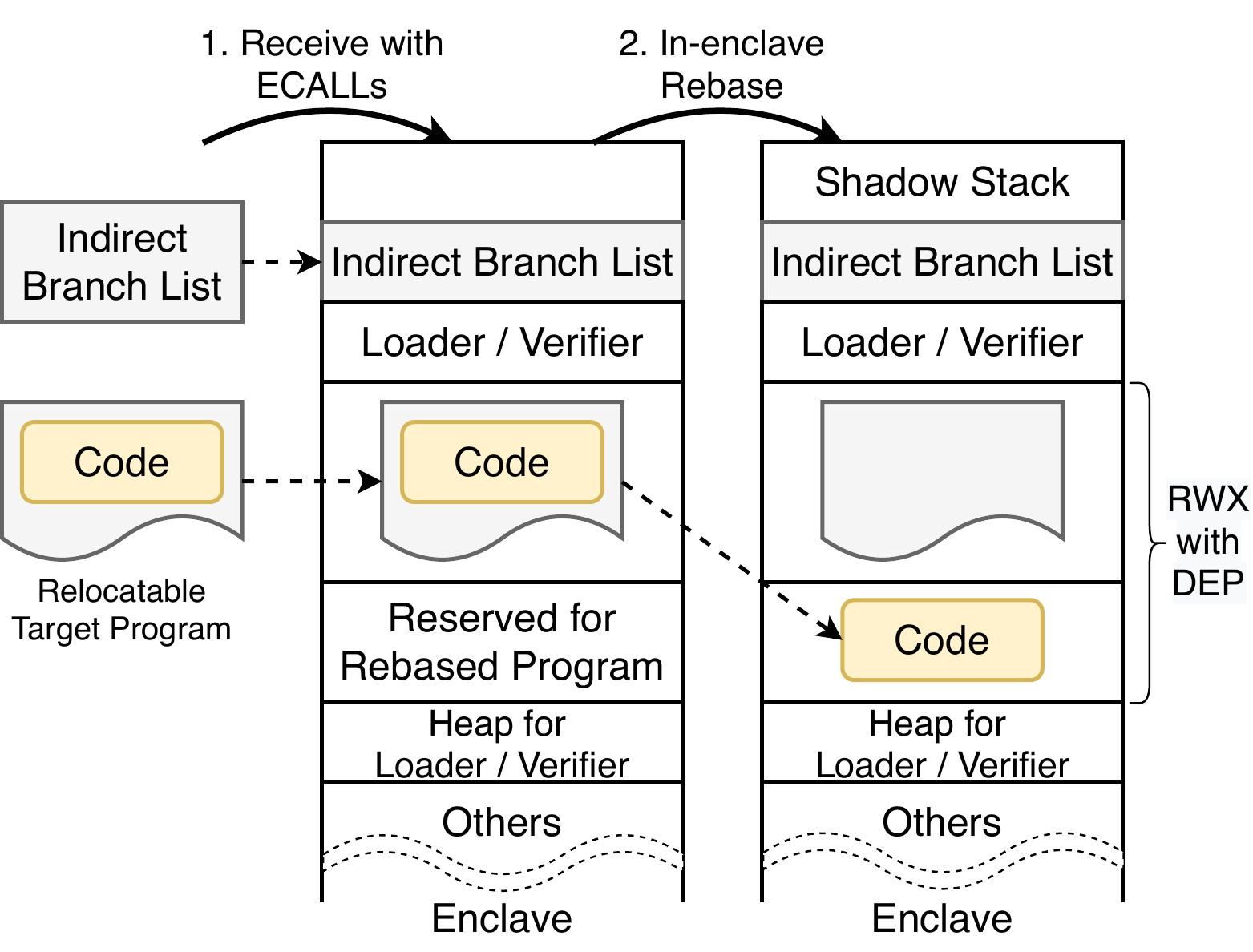}}
\caption{Detailed workflow of the dynamic loader}\label{fg-dynloader}
\label{fig}
\end{figure}

\vspace{3pt}\noindent\textbf{Remote attestation}.\label{subsec:ra-impl}
Once the bootstrap enclave is initiated, it needs to be attested. We leverage the original RA routine~\cite{originalra} and adjust it to our design. The original RA routine requires that the host, which is assumed to run the enclave as the `client', initiates the attestation towards the `server', who owns the data. While in this CCaaS scenario, the service runs in the enclave while the remote user owns the data. So, we modify this routine to enable a remote CCaaS user to initiate the attestation.

The RA procedures can be invoked by calling \verb|sgx_ra_init()| inside the service provider’s enclave after secret provision between the remote user and the service provider. After obtaining an enclave quote of the bootstrap enclave which is signed with the platform's EPID key, the remote data owner can submit the quote to IAS and obtain an attestation report. 


\vspace{3pt}\noindent\textbf{Dynamic loader}. When the RA is finished, the trust between data owner and the bootstrap enclave is established. Then the user can locally/remotely call the Ecall (\verb|ecall_receive_binary|) to load the service binary instrumented with security annotations and the indirect branch list without knowing the code. 
User data is loaded from untrusted memory into the trusted enclave memory when the user remotely calls Ecall (\verb|ecall_receive_userdata|), to copy the data to the section reserved for it.

Then, the dynamic loader in the bootstrap enclave loads and relocates the generated code. 
The indirect branch list, which is comprised of symbol names that will be checked in indirect branch instrumentations, will be resolved at the very beginning. 
In our implementation, there are both 4M memory space for storing indirect branch targets, as well as for shadow stack.
And we reserve 64M memory space for received binary and for `.data' section. The heap base address is slightly larger than the end of received binary, and 0x27000 Bytes (156 KB) space is reserved for the loader's own heap. After relocation, the detailed memory layout and some key steps are shown in Figure~\ref{fg-dynloader}.

\vspace{3pt}\noindent\textbf{Policy verifier}.\label{subsec-boundarychecking}
The policy-compliance verifier, is composed with three components - a clipped disassembler, a verifier, and a immediate operand rewritter.

\vspace{2pt}\noindent$\bullet$\textit{ Clipped disassembler.} 
We enforce each policy mostly at assembly level. 
Thus, we incorporate a lightweight disassembler inside the enclave. To implement the disassembler, we remove unused components of this existing wide-used framework, and use Recursive Descent Disassembly to traverse the code. We used the \textit{diet} mode, in which some non-critical data are removed, thus making the engine size at least 40\% smaller~\cite{quynh2014capstone}. 

\vspace{2pt}\noindent$\bullet$\textit{ Policy verifier.}\label{subsec-policyverifer}
The verifier and the following rewriter do the work just right after the target binary is disassembled, according the structured guard formats provided by our code generator. The verifier uses a simple scanning algorithm to ensure the policies applied in assembly language instrumentation. 
Specifically, the verifier scans the whole assembly recursively along with the disassembler. It follows the clipped disassembler to scan instrumentations before/after certain instructions are in place, and checks if there is any branch target pointing between instructions in those instrumentations.

\vspace{2pt}\noindent$\bullet$\textit{ Imm rewriter.}\label{subsec:immrewriter} One last but not least step before executing the target binary code is to resolve and replace the Imm operands in instrumentations, including the base of the shadow stack, and the addresses of indirect branch targets (i.e. legal jump addresses). For example, the genuine base address of shadow stack is the start address \verb|__ss_start| of the memory space reserved by the bootstrap enclave for the shadow stack. And the ranges are determined using functions of Intel SGX SDK during dynamic loading (Section~\ref{subsec:verify}).

We use the simplest way to rewrite Imm operands. Table~\ref{tb-rewritter} shows what the specific values should be before and after rewriting, respectively. 
The first column of table~\ref{tb-rewritter} shows the target we need to rewrite while loading. For instance, the upper bound address of data section would be decided during loading, but it would be 3ffffffffffffffff (shown in the 2nd. column) during the proof generation and will be modified to the real upper data bound address. The third column shows the variable name used in our prototype.

\begin{table}[htbp]
\footnotesize
\caption{Fields to be rewritten}\label{tb-rewritter}
\begin{center}
\begin{tabular}{|c|c|c|}
\hline
\textbf{Target imm description} & \textbf{From} & \textbf{To} \\
\hline
Upper bound of data section & 3ffffffffffffffff & \verb|upper_data_bound| \\
\hline
Lower bound of data section & 4ffffffffffffffff & \verb|lower_data_bound| \\
\hline
Upper bound of stack & 5ffffffffffffffff & \verb|upper_stack_bound| \\
\hline
Lower bound of stack & 6ffffffffffffffff & \verb|lower_stack_bound| \\
\hline
Upper bound of code section & 7ffffffffffffffff & \verb|lower_code_bound| \\
\hline
Lower bound of code section & 8ffffffffffffffff & \verb|lower_code_bound| \\
\hline 
\hline 
\# of indirect branch targets & 1ffffffff & \verb|branch_target_idx| \\
\hline
Addr. of branch target list & 1ffffffffffffffff & \verb|__branch_target| \\
\hline
Addr. of the shadow stack & 2ffffffffffffffff & \verb|__ss_start| \\
\hline
\end{tabular}
\end{center}
\end{table}

\section{Evaluation}\label{sec-evaluation}
In this section we report our security analysis and performance evaluation of CAT-SGX.
\subsection{Security Analysis}\label{subsec-securityanalysis}

\noindent\textbf{TCB analysis}. 
The hardware TCB of CAT-SGX includes the TEE-enabled platform, i.e. the SGX hardware. The software TCB includes the following components to build the bootstrap enclave.

\vspace{2pt}\noindent$\bullet$\textit{ Loader and verifier}.
The loader we implemented consists of
less than 600 lines of code (LoCs) and the verifier includes less than 700 LoCs, which also integrates SGX SDK and part of Capstone libraries. 

\vspace{2pt}\noindent$\bullet$\textit{ ECall/OCall stubs for supporting P0}. This was implemented in less than 500 LoCs. 

\vspace{2pt}\noindent$\bullet$\textit{ Simple RA protocol realization}. The implementation  (Section~\ref{subsec:ra-impl}) introduces about 200 LoCs.

\noindent Altogether, our software TCB contains less than 2000 LoCs and some dependencies, which was compiled into a self-contained binary with 1.9 MB in total.


\vspace{3pt}\noindent\textbf{Policy analysis}. 
Here we show how the policies on the untrusted code, once enforced, prevent information leaks from the enclaves. In addition to side channels, there are two possible ways for a data operation to go across the enclave boundaries:
bridge functions~\cite{van2019tale} and memory write.

\vspace{2pt}\noindent$\bullet$\textit{ Bridge functions}. 
With the enforcement of P0, the loaded code can only invoke our OCall stubs, which prevents the leak of plaintext data through encryption and controls the amount of information that can be sent out (to the code provider in CDaaS).

\vspace{2pt}\noindent$\bullet$\textit{ Memory write operations}. All memory writes, both direct memory store and indirect register spill, are detected and blocked. Additionally, software DEP is deployed so the code cannot change itself.  Also the control-flow integrity (CFI) policy, P5, prevents the attacker from bypassing the checker with carefully constructed gadgets by limiting the control flow to only legitimate target addresses. 

As such, possible ways of information leak to the outside of the enclave are controlled. As proved by previous works~\cite{sinha2015moat,sinha2016design} the above-mentioned policies (P1\textasciitilde P5) guarantee the property of confidentiality. Furthermore the policy (P5) of \textit{protecting return addresses and indirect control flow transfer, together with preventing writes to outside} has been proved to be adequate to construct the confinement~\cite{schuster2015vc3,sinha2016design}. So, enforcement of the whole set of policies from P0 to P5 is sound and complete in preventing explicit information leaks. 
In the meantime, our current design is limited in side-channel protection. We can mitigate the threats of page-fault based attacks and exploits on L1/L2 cache once Hyper-threading is turned off or HyperRace~\cite{chen2018racing} is incorporated (P6). However, defeating the attacks without triggering interrupts, such as inference through LLC is left for future research.

\subsection{Performance Evaluation}\label{subsec-experiments}




\begin{table*}[htbp]
\footnotesize
\caption{Binary code size and execution time of simple applications}\label{tb-simple-perf}
\begin{center}
\begin{tabular}{|c|c|c|c|c|c|c|c|}
\hline
\textbf{Application Name} & \textbf{Size} & \textbf{Size (P1\textasciitilde P5)} & \textbf{Size (P1\textasciitilde P6)} & \textbf{Execution Time} & \textbf{Execution Time (P1\textasciitilde P5)} & \textbf{Execution Time (P1\textasciitilde P6)}\\
\hline
bm\_clock & 209KB & 217KB (+3.83\%) &  218KB (+4.31\%) & 1.271s & 1.457s (+14.6\%) & 1.469s (+15.8\%)\\
\hline
bm\_malloc\_and\_magic & 227KB & 237KB (+4.41\%) &  239KB (+5.29\%) & 1.343s & 1.537s (+14.4\%)  & 1.638s (+22.0\%)\\
\hline
bm\_malloc\_memalign & 229KB & 240KB (+4.80\%)  & 242KB (+5.68\%) & 1.278s & 1.467s (+14.8\%)  & 1.567s (+22.6\%)\\
\hline
bm\_malloc\_and\_sort & 208KB & 222KB (+6.73\%) & 225KB (+8.17\%) & 1.270s & 1.473s (+16.0\%) & 1.620s (+27.6\%)\\
\hline
bm\_memcpy & 4.7KB & 7.9KB (+68.1\%) &  11KB (+134\%) & 1.211s & 1.247s (+2.97\%) & 1.396s (+15.3\%)\\
\hline
bm\_memchr & 5.2KB & 8.3KB (+59.6\%) & 11KB (+116\%) & 1.210s & 1.251s (+3.39\%) & 1.391s (+15.0\%)\\
\hline
bm\_sprintf & 70KB & 74KB (+5.71\%) & 76KB (+8.57\%) & 1.218s & 1.299s (+6.65\%) & 1.440s (+18.2\%)\\
\hline
bm\_sort\_and\_binsearch & 89KB & 98KB (+10.1\%) & 102KB (+14.6\%) & 1.234s & 1.314s (+6.48\%) & 1.460s (+18.3\%)\\
\hline 
\end{tabular}
\end{center}
\end{table*}

\begin{table}[htbp]
\footnotesize
\caption{Performance evaluation on nBench}\label{tb-nben-perf}
\begin{center}
\begin{tabular}{|c|c|c|c|}
\hline
\textbf{Program Name} & \textbf{Baseline} & \textbf{P1\textasciitilde P5} & \textbf{P1\textasciitilde P6}\\
\hline
NUMERIC SORT & 1487$\mu$s  & 1588$\mu$s (+6.79\%) & 1665$\mu$s (+12.0\%) \\
\hline
STRING SORT & 8460$\mu$s & 9507$\mu$s (+12.4\%)  & 10.02ms (+18.4\%)\\
\hline
BITFIELD & 46.83ns & 54.10ns (+15.5\%)  & 55.23ns (+17.9\%) \\
\hline
FP EMULATION & 14.93ms & 14.98ms (+0.33\%) & 15.73ms (+5.36\%) \\
\hline
FOURIER & 34.22$\mu$s & 35.21$\mu$s (+2.89\%) & 36.77$\mu$s (+7.45\%) \\
\hline
ASSIGNMENT & 43.52ms & 54.41ms (+25.0\%)  & 60.85ms (+39.8\%)\\
\hline
IDEA & 342.2$\mu$s & 352.9$\mu$s (+3.13\%)  & 385.7$\mu$s (+12.1\%) \\
\hline
HUFFMAN & 550.1$\mu$s & 649.6$\mu$s (+18.1\%)  & 667.1$\mu$s (+21.3\%)\\
\hline
NEURAL NET & 49.44ms  & 59.43ms (+20.2\%) & 60.84ms (+23.1\%) \\
\hline
LU DECOMPOSITION & 1024$\mu$s & 1123$\mu$s (+9.67\%)  & 1255$\mu$s (+22.6\%)\\
\hline
\end{tabular}
\end{center}
\end{table}

\noindent\textbf{Testbed setup}. 
In our research, we evaluated the performance of our prototype and tested its code generation and code execution. All experiments were conducted on Ubuntu 18.04 (Linux kernel version 4.4.0) with SGX SDK 2.5 installed on Intel Xeon CPU E3-1280 with 64GB memory. 
Also we utilized  GCC 5.4.0 to build the bootstrap enclave and the SGX application, and the parameters `-fPIC', `-fno-asynchronous-unwind-tables', `-fno-addrsig', and `-mstackrealign' to generate X86 binaries. 

\vspace{3pt}\noindent\textbf{Performance on simple applications}. We used the applications provided by the SGX-Shield project~\cite{seo2017sgx} as a micro-benchmark. In our experiment, we ran each test case for 10 times, measured the resource consumption in each case and reported the median value. Specifically, we first set the baseline as the performance of an uninstrumented program running through a pure loader (a loader that only does the dynamic loading but no policy-compliance checking). The we compared the baseline with the performance of instrumented programs to measure the overheads. Also the compilation time of each micro-benchmark varies from several seconds to tens of seconds, which is negligible compared with conventional PCC methods (2\textasciitilde 5$\times$)~\cite{necula2001oracle}.


Table~\ref{tb-simple-perf} illustrates overheads of our approach. 
From the table, we can see that the size of instrumented binaries (aka. the ``code + proof'') is 18.1\% larger than the original code and their executions were delayed by 9.8\% on average
when only P1\textasciitilde P5 are enforced. It becomes 130\% in memory and 119\% in time when all policies, including P6, are enforced. Note that this batch of benchmarks are mostly a `first-simple-processing-then-syscall' program. At the worst case - `bm\_malloc\_and\_sort', CAT-SGX showed 27.6\% overhead in execution time. 

\vspace{3pt}\noindent\textbf{Performance on nBench}. We instrumented all applications in the SGX-nBench~\cite{sgxnbench}, and ran each testcase of the nBench suites under a few settings, each for 10 times. These settings include just explicit memory write check (P1), both explicit memory write check and implicit stack write check (P1+P2), all memory write and indirect branch check (P1\textasciitilde P5), and together with side channel mitigation (P1\textasciitilde P6). 

Table~\ref{tb-nben-perf} shows the average execution time under different settings.
Without side channel mitigation (P1\textasciitilde P5), CAT-SGX introduces an 0.3\% to 25\% overhead (on FP-emulation). 
Apparently, the store instruction instrumentation alone (P1) does not cause a large performance overhead, with largest being 6.7\%. Also, when P1 and P2 are applied together, the overhead just becomes slightly higher than P1 is enforced alone. 
Besides, almost all benchmarks in nBench perform well under the CFI check P5 (less than 3\%) except for the benchmarks Bitfield (whose overhead is about 4\%) and the Assignment (about 10\% due to its frequent memory access pattern).

\vspace{3pt}\noindent\textbf{Performance on real-world applications}.  We further evaluated our prototype on various real-world applications, including personal health data analysis, personal financial data analysis, and Web servers. 
We implemented those macro-benchmarks and measured the differences between their baseline performance (without instrumentation) in enclave and the performance of our prototype.

\vspace{2pt}\noindent$\bullet$\textit{ Sensitive health data analysis}.  We studied the following two applications: 

\noindent 1) Sequence Alignment. We implemented the Needleman–Wunsch algorithm~\cite{needleman1970general} that aligns two human genomic sequences in the FASTA format~\cite{fasta-format} taken from the 1000 Genomes project~\cite{1000genomes}. The algorithm uses dynamic programming to compute recursively a two dimensional matrix of similarity scores between subsequences; as a result, it takes $N^2$ memory space where $N$ is the length of the two input sequences.

Again, we measured the program execution time under the aforementioned settings. Figure~\ref{fg-nw-perf} shows the performance of the algorithm with different input lengths (x-axis). The overall overhead (including all kinds of instrumentations) is no more than 20\% (with the P1 alone no more than 10\%), when input size is small (less than 200 Bytes). When input size is greater than 500 Bytes, the overhead of P1+P2 is about 19.7\%  while P1\textasciitilde P5 spends 22.2\% more time than the baseline.

\begin{figure}[htbp]
\centerline{\includegraphics[scale=0.48]{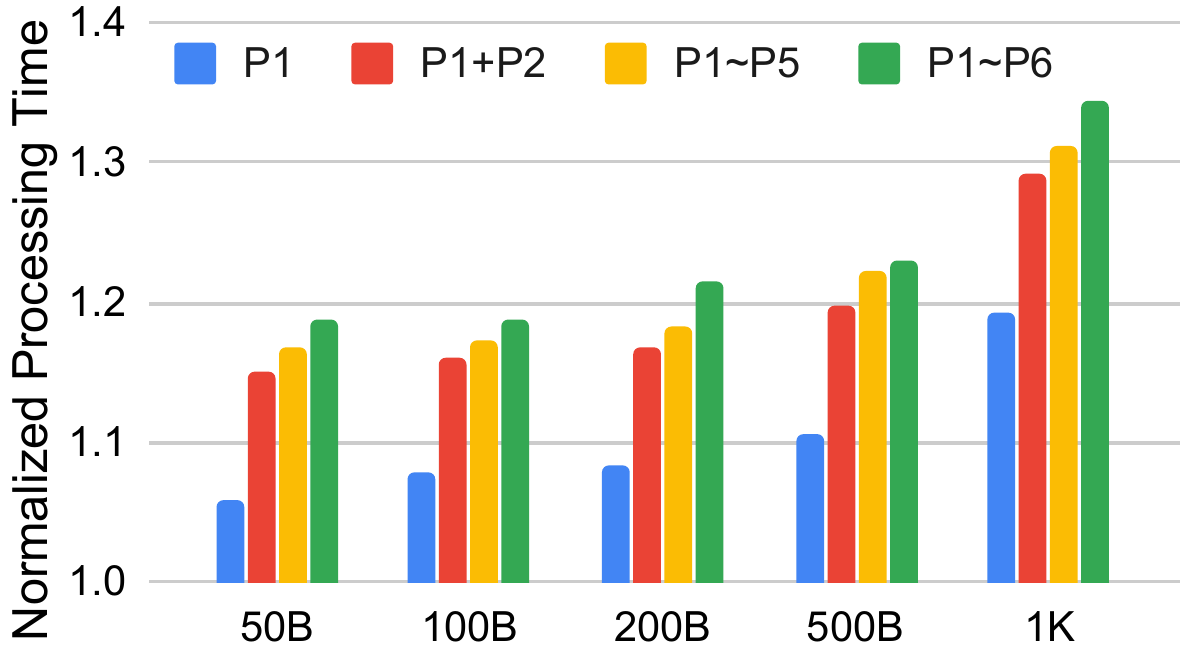}}
\caption{Performance on sequence alignment}\label{fg-nw-perf}
\end{figure}

\begin{figure}[htbp]
\centerline{\includegraphics[scale=0.48]{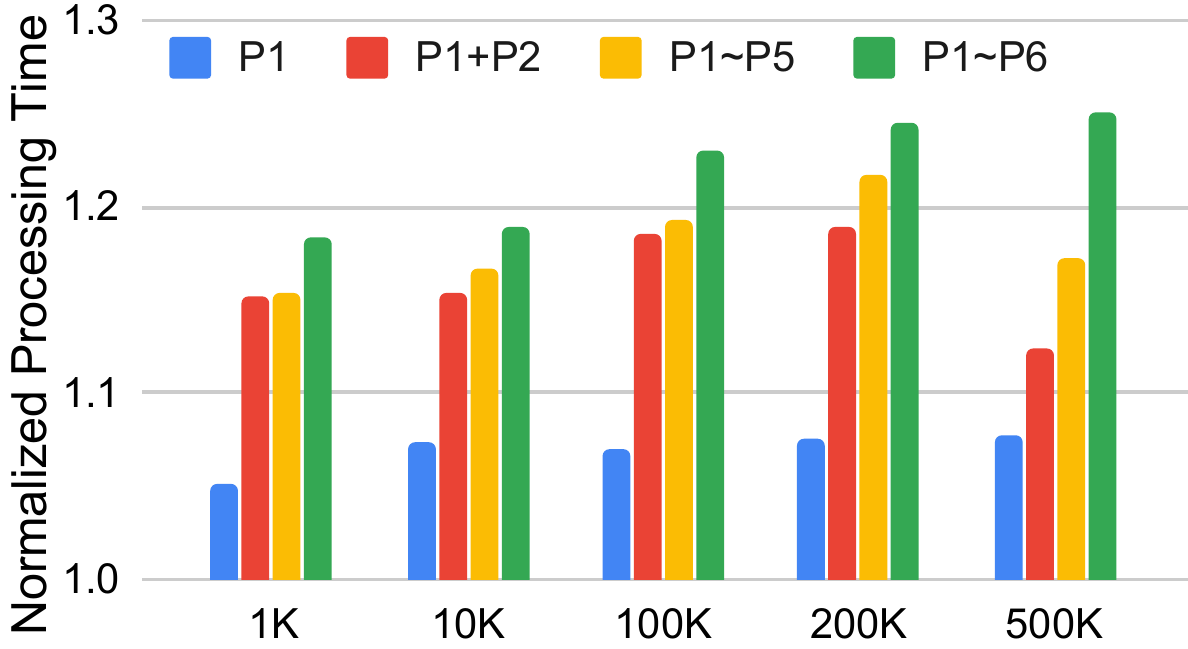}}
\caption{Performance on sequence generation}\label{fg-fasta-perf}
\end{figure}
    
\noindent 2) Sequence Generation. We re-implemented the FASTA benchmark~\cite{fasta}, which is designed to simulate DNA sequences based on pre-defined nucleotide frequencies. The program can output the nucleotide sequences of length 2$N$, 3$N$ and 5$N$, where $N$ is used to measure the output size. Figure~\ref{fg-fasta-perf} shows the performance when the output size (x-axis) varies from 1K to 500K nucleotides. Enforcing P1 alone results in 5.1\% and 6.9\% overheads when 1K and 100K are set as the output lengths. When the output size is 200K, our prototype yields less than 20\% overhead. Even when the side channel mitigation is applied, the overhead becomes just 25\%. With the increase of processing data size, the overhead of the system also escalates; however, the overall performance remains acceptable.

\vspace{2pt}\noindent$\bullet$\textit{ Personal credit score analysis}. We further looked into another realistic and more complicated workload. Credit scoring is a typical application that needs to be protected more carefully - both the credit card holder's cost calendar and card issuer's scoring algorithm need to be kept secret. In our study, we implemented a BP neural network-based credit scoring algorithm~\cite{jensen1992using} that calculates user's credit scores. The input file contains users' history records and the output is a prediction whether the bank should approve the next transaction. The model was trained on 10000 records and then used to make prediction (i.e., output a confidence probability) on different test cases.

\begin{figure}[htbp]
\centerline{\includegraphics[scale=0.48]{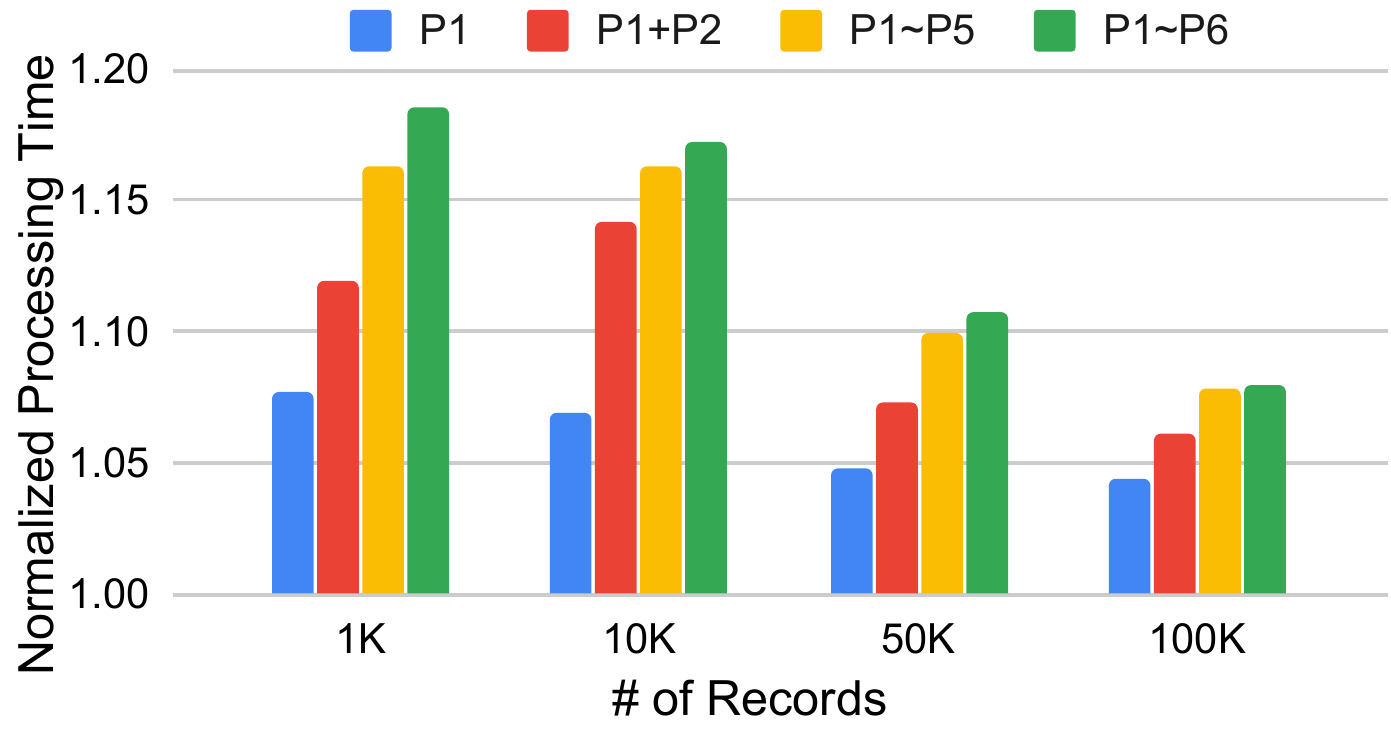}}
\caption{Performance on credit scoring}\label{fg-credit-score}
\end{figure}

As shown in Figure~\ref{fg-credit-score}, on 1000 and 10000 records, enforcement of P1\textasciitilde P5 would yields around 15\% overhead. 
While processing more than 50000 records, the overhead of the full check does not exceed 20\%.  The overhead of P1\textasciitilde P6 does not exceed 10\% when processing 100K records.

\begin{figure}[htbp]
\centerline{\includegraphics[scale=0.48]{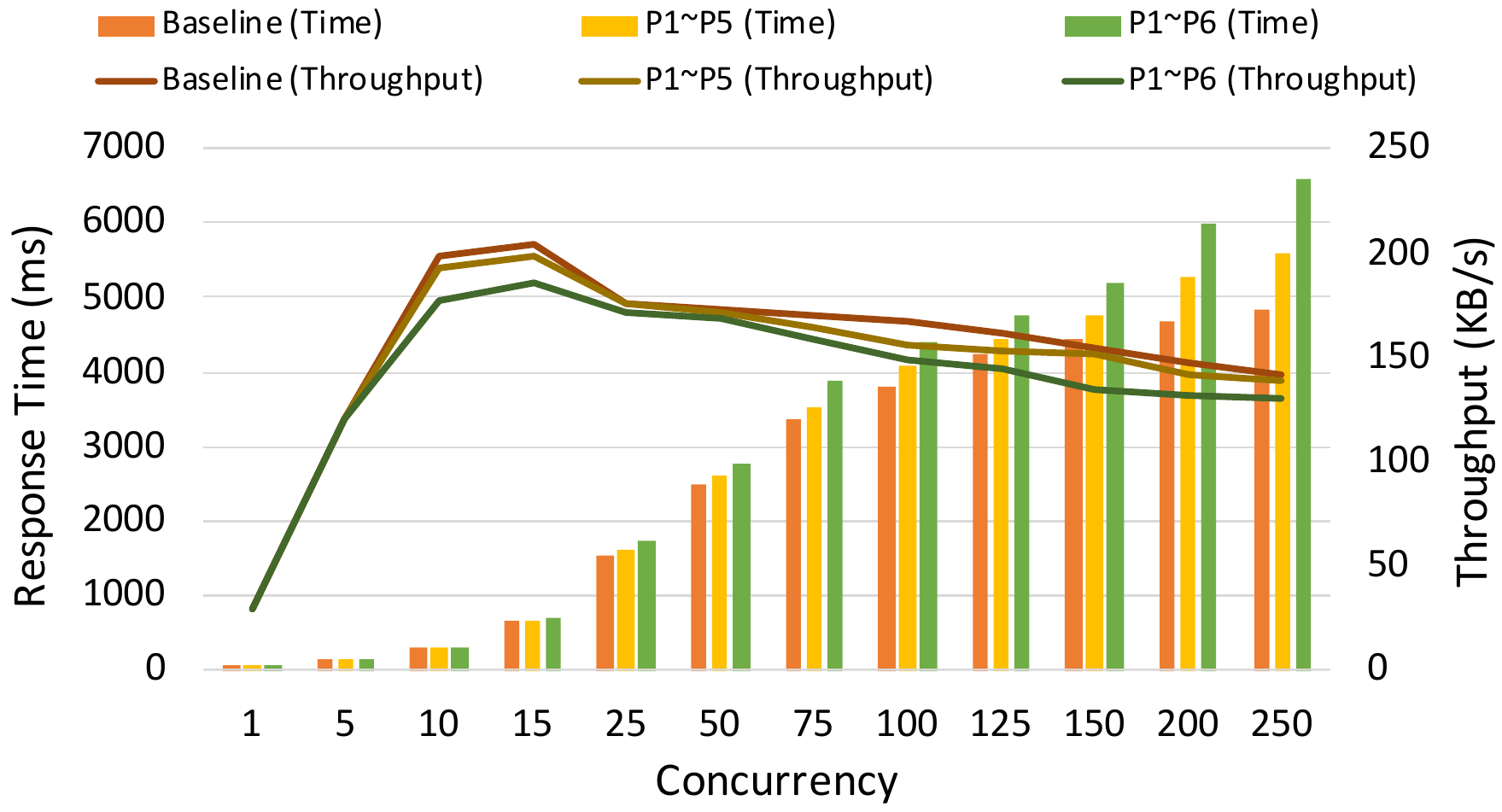}}
\caption{Performance on HTTPS server}\label{fg-https-all}
\end{figure}

\vspace{2pt}\noindent$\bullet$\textit{ HTTPS server}. We also built an HTTPS server to run in enclave using the mbed TLS library~\cite{mbedtls}. Our protection only allows two system calls (\texttt{send/recv}) to be executed via the OCall stubs for client/server communication. A client executes a stress test tool - Siege~\cite{siege} - on another host in an isolated LAN. Siege was configured to send continuous HTTPS requests (with no delay between two consecutive ones) to the web server for 10 minutes. We measured its performance in the presence of different concurrent connections to understand how our instrumented HTTPS server implementation would perform. 


Figure~\ref{fg-https-all} shows the response times and throughput when all policies are applied to the HTTPS server benchmark. When the concurrent connections are less than 75, the instrumented HTTPS server has similar performance of the in-enclave https server without instrumentation. When the concurrency increases to 100, the performance goes down to some extent. 
While after the concurrency increases to 150, the response time of instrumented server goes up significantly. On average, enforcing P1\textasciitilde P6 results in 14.1\% overhead in the response time. As for throughput, when the number of the concurrent connections is between 75 and 200, the overhead is less than 10\%. 
These experiments on realistic workloads show that all policies, including side-channel mitigation, can be enforced at only reasonable cost. 



\ignore{
In this section we report both security analysis and performance evaluation of the proposed CAT-SGX scheme.
\subsection{Security Evaluation}\label{subsec-securityanalysis}

\noindent\textbf{TCB analysis}. 
In accordance with the design, the hardware TCB includes the TEE-enabled platform, i.e. the SGX hardware. The software TCB includes the following components to build the bootstrap enclave.

\vspace{2pt}\noindent$\bullet$\textit{ Loader and verifier}.
The loader we implemented consists of
less than 600 lines of code (LoCs) which is made from scratch, and the verifier consists of less than 700 LoCs (supported by SGX SDK and Capstone, counted below).

\vspace{2pt}\noindent$\bullet$\textit{ ECall/OCall stubs for supporting P0}. This is implemented in less than 500 LoCs. 

\vspace{2pt}\noindent$\bullet$\textit{ Simple RA protocol realization}. The implementation  (Section~\ref{subsec:ra-impl}) introduces about 200 LoCs.

\vspace{2pt}\noindent$\bullet$\textit{ Dependencies}. Our framework is built upon several existing libraries, which include the SGX SDK libraries, parts of Capstone libraries, and an ELF parser.



\vspace{3pt}\noindent\textbf{Policy justification}. 
Now we study how the enforced policies on the untrusted code prevent information leakages from the enclaves.
Except for side channels, there are two possible way for data operations crossing the enclave boundaries:
bridge functions~\cite{van2019tale} and memory write operations.

\vspace{2pt}\noindent$\bullet$\textit{ Bridge functions}. 
With the enforcement of P0, the loaded code can only invoke the customized OCall stubs, which prevents the leakage of plaintext data with encryption and controls the amount of information that can be output.

\vspace{2pt}\noindent$\bullet$\textit{ Memory write operations}. All memory writes, both direct memory stores and indirect register spill, can be detected and blocked. Additionally, software DEP is deployed so the code cannot change itself.  CFI policies are enforced to prevent the attacker from bypassing the checker with carefully constructed gadgets.

As such, possible ways of leakage to the outside of the enclave are screened. As proved by previous works~\cite{sinha2015moat,sinha2016design} the above-mentioned policies (P1\textasciitilde P5) guarantee the property of confidentiality. Furthermore the policies of \textit{protecting return addresses}, \textit{protecting indirect control flow transfers}, and \textit{preventing writes to outside} have been proved adequate to construct the confinement~\cite{schuster2015vc3,sinha2016design}. In conclusion enforcing the whole set of policies from P0 to P6 is sound and complete in verifying confidentiality.

\subsection{Performance Evaluation}\label{subsec-experiments}


We now focus on the overhead evaluation. The benchmarks include various workloads. We quantify the time and space costs of our prototype by enforcing the execution of the described policies.

\begin{table*}[htbp]
\footnotesize
\caption{Binary code size and execution time of simple applications}\label{tb-simple-perf}
\begin{center}
\begin{tabular}{|c|c|c|c|c|c|c|c|}
\hline
\textbf{Application Name} & \textbf{Size} & \textbf{Size (P1\textasciitilde P5)} & \textbf{Size (P1\textasciitilde P6)} & \textbf{Execution Time} & \textbf{Execution Time (P1\textasciitilde P5)} & \textbf{Execution Time (P1\textasciitilde P6)}\\
\hline
bm\_clock & 209KB & 217KB (+3.83\%) &  218KB (+4.31\%) & 1.271s & 1.457s (+14.6\%) & 1.469s (+15.8\%)\\
\hline
bm\_malloc\_and\_magic & 227KB & 237KB (+4.41\%) &  239KB (+5.29\%) & 1.343s & 1.537s (+14.4\%)  & 1.638s (+22.0\%)\\
\hline
bm\_malloc\_memalign & 229KB & 240KB (+4.80\%)  & 242KB (+5.68\%) & 1.278s & 1.467s (+14.8\%)  & 1.567s (+22.6\%)\\
\hline
bm\_malloc\_and\_sort & 208KB & 222KB (+6.73\%) & 225KB (+8.17\%) & 1.270s & 1.473s (+16.0\%) & 1.620s (+27.6\%)\\
\hline
bm\_memcpy & 4.7KB & 7.9KB (+68.1\%) &  11KB (+134\%) & 1.211s & 1.247s (+2.97\%) & 1.396s (+15.3\%)\\
\hline
bm\_memchr & 5.2KB & 8.3KB (+59.6\%) & 11KB (+116\%) & 1.210s & 1.251s (+3.39\%) & 1.391s (+15.0\%)\\
\hline
bm\_sprintf & 70KB & 74KB (+5.71\%) & 76KB (+8.57\%) & 1.218s & 1.299s (+6.65\%) & 1.440s (+18.2\%)\\
\hline
bm\_sort\_and\_binsearch & 89KB & 98KB (+10.1\%) & 102KB (+14.6\%) & 1.234s & 1.314s (+6.48\%) & 1.460s (+18.3\%)\\
\hline 
\end{tabular}
\end{center}
\end{table*}

\begin{table}[htbp]
\footnotesize
\caption{Performance evaluation on nBench}\label{tb-nben-perf}
\begin{center}
\begin{tabular}{|c|c|c|c|}
\hline
\textbf{Program Name} & \textbf{Baseline} & \textbf{P1\textasciitilde P5} & \textbf{P1\textasciitilde P6}\\
\hline
NUMERIC SORT & 1487$\mu$s  & 1588$\mu$s (+6.79\%) & 1665$\mu$s (+12.0\%) \\
\hline
STRING SORT & 8460$\mu$s & 9507$\mu$s (+12.4\%)  & 10.02ms (+18.4\%)\\
\hline
BITFIELD & 46.83ns & 54.10ns (+15.5\%)  & 55.23ns (+17.9\%) \\
\hline
FP EMULATION & 14.93ms & 14.98ms (+0.33\%) & 15.73ms (+5.36\%) \\
\hline
FOURIER & 34.22$\mu$s & 35.21$\mu$s (+2.89\%) & 36.77$\mu$s (+7.45\%) \\
\hline
ASSIGNMENT & 43.52ms & 54.41ms (+25.0\%)  & 60.85ms (+39.8\%)\\
\hline
IDEA & 342.2$\mu$s & 352.9$\mu$s (+3.13\%)  & 385.7$\mu$s (+12.1\%) \\
\hline
HUFFMAN & 550.1$\mu$s & 649.6$\mu$s (+18.1\%)  & 667.1$\mu$s (+21.3\%)\\
\hline
NEURAL NET & 49.44ms  & 59.43ms (+20.2\%) & 60.84ms (+23.1\%) \\
\hline
LU DECOMPOSITION & 1024$\mu$s & 1123$\mu$s (+9.67\%)  & 1255$\mu$s (+22.6\%)\\
\hline
\end{tabular}
\end{center}
\end{table}

\vspace{3pt}\noindent\textbf{Testbed setup.} As same as the implementation, our evaluation was also completed on Linux/X86. We tested our prototype's performance on code generation and code execution. All experiments were conducted on Ubuntu 18.04 (Linux kernel version 4.4.0) with SGX SDK 2.5 installed on Intel Xeon CPU E3-1280 with 64GB memory. 
And we use GCC 5.4.0 to build the bootstrap enclave and the SGX application. For code generationr, we use `-fPIC', `-fno-asynchronous-unwind-tables', `-fno-addrsig', and `-mstackrealign' as parameters to generate X86 binaries. 

\vspace{3pt}\noindent\textbf{Performance on simple applications.} We reused and adapted the applications provided by SGX-Shield~\cite{seo2017sgx} as the simplest benchmarks. We run each test case of this micro benchmark for 10 times and report the median value. 
Specifically, we first measured the runtime overhead with the baseline being an uninstrumented program running in a pure loader (a loader that only does the dynamic loading but no policy-compliance checking). Then we instrument those programs and measure the execution time.
The compilation time of each micro-benchmark varies from several seconds to tens of seconds, which is negligible compared with conventional PCC methods (2\textasciitilde 5$\times$)~\cite{necula2001oracle}.

Table~\ref{tb-simple-perf} indicates the execution time of native and instrumented code. 
From it we can see that the size of instrumented binaries (aka. the ``code + proof'') are 3.8\% - 68.1\% larger than the original code if we only enforce P1\textasciitilde P5.
This batch of benchmarks are mostly a `first-simple-processing-then-syscall' program. At the worst case - `bm\_malloc\_and\_sort', CAT-SGX showed 38.3\% overhead. 

\vspace{3pt}\noindent\textbf{Performance on nBench.} We instrumented all applications in the SGX-nBench~\cite{sgxnbench}, and run each testcase of the nBench suites in different instrumentation levels for 10 times. We performed the measurements at different granularities - just explicit memory write check (P1), both explicit memory write check and implicit stack  write check (P1+P2), all memory write and indirect branch check (P1\textasciitilde P5), and checks with alternative side channel mitigations (P1\textasciitilde P6). 

Table~\ref{tb-nben-perf} indicates the average execution time of different instrumentation levels.
Once protections without side channel defenses (P1\textasciitilde P5) are applied, CAT-SGX imposes 0.3\% (on FP-emulation) and more overhead. The Assignment algorithm - a well-known task allocation algorithm - takes the largest overhead, while just P1 would increase 6.7\% execution time. Certainly, this store instruction instrumentation alone does not cause a large performance loss. Also, when P1 and P2 applied, the overhead just becomes slightly higher than P1 was applied. 
Besides, almost all benchmarks in nBench suffer few overhead on CFI check (less than 3\%) except for the benchmarks Bitfield (which costs about 4\%) and the Assignment (which costs nearly 10\% due to its frequent memory access pattern).

\vspace{3pt}\noindent\textbf{Performance on real-world cases.} We now turn to multiple realistic cases on personal health data analysis, personal financial data analysis, and Web servers. 
We implement those macro-benchmarks and evaluate the differences between their baseline performance running in enclave and the performance on our prototype.

\vspace{2pt}\noindent$\bullet$\textit{ Sensitive health data analysis.}

1) Sequence Alignment. We implemented the Needleman–Wunsch algorithm~\cite{needleman1970general} that aligns two human genomic sequences in FASTA format~\cite{fasta-format} taken from the 1000 Genomes project ~\cite{1000genomes}. The algorithm applied the dynamic programming to compute recursively a two dimensional matrix of similarity scores between subsequences; as a result, it takes $N^2$ memory space where $N$ is the length of the two input sequences are.

Still, we measured the program execution time at different granularities. Figure~\ref{fg-nw-perf} shows the performance of the algorithm for different sizes of inputs (x-axis). The overall performance overhead (including all kinds of instrumentations) is no more than 20\% (with the P1 overhead no more than 10\%), when input size is small (less than 200 Bytes). When input size is greater than 500 Bytes, the overhead of P1+P2 is about 19.7\% overhead while P1\textasciitilde P5 spends 22.2\% more time than the baseline.

\begin{figure}[htbp]
\centerline{\includegraphics[scale=0.48]{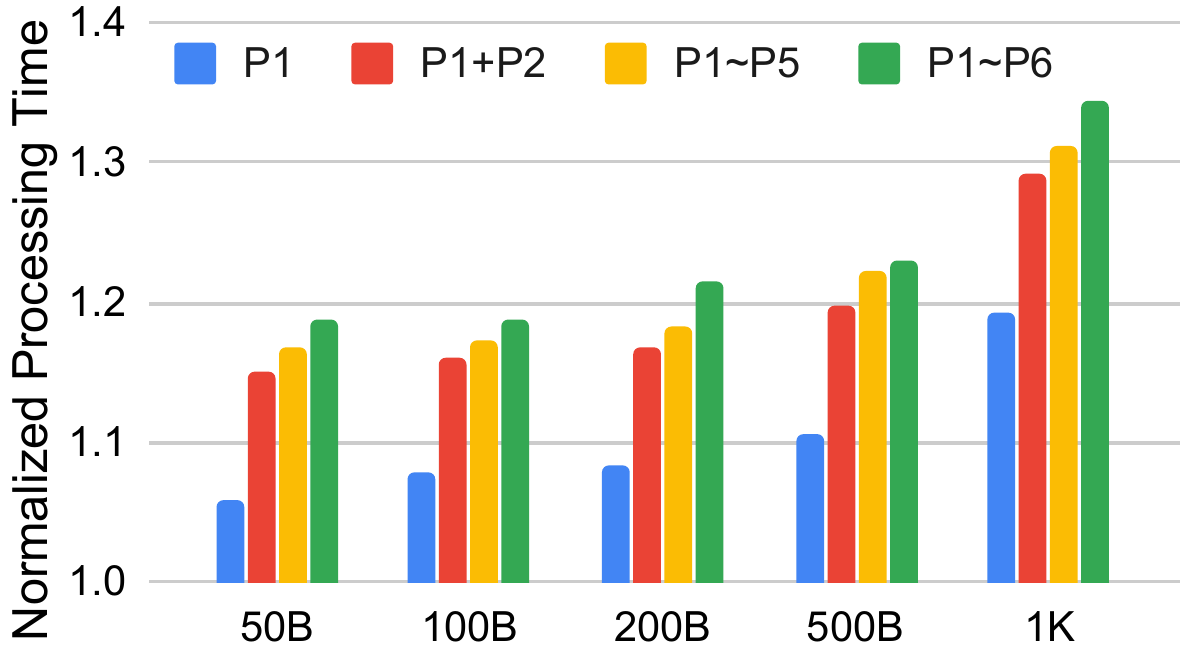}}
\caption{Performance on sequence alignment}\label{fg-nw-perf}
\end{figure}

\begin{figure}[htbp]
\centerline{\includegraphics[scale=0.48]{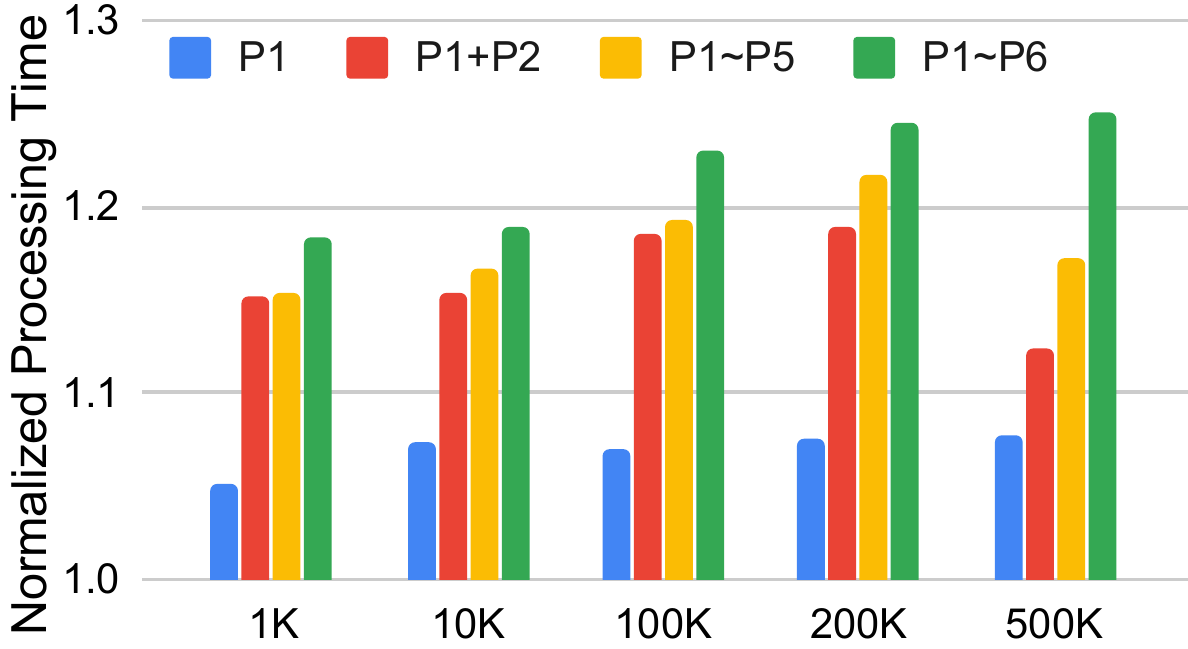}}
\caption{Performance on sequence generation}\label{fg-fasta-perf}
\end{figure}
    
2) Sequence Generation. We re-implemented the FASTA benchmark~\cite{fasta}, which is designed to simulate DNA sequences based on pre-defined nucleotide frequencies. The program can output the nucleotide sequences of length 2$N$, 3$N$ and 5$N$, where $N$ is used to measure the output size. Figure~\ref{fg-fasta-perf} displays the performance when the output size (x-axis) varies from 1K to 500K nucleotides. Enforcing P1 alone results in 5.1\% and 6.9\% greater overhead when 1K and 100K were designated as the output length. When the output size is 200K, our prototype yields less than 20\% overhead. Even when alternative side channel mitigation techniques are applied, the overhead would exceed 125\%. With the increase of processing data size, the performance overhead of the system also escalates; however, the overall performance remains acceptable.

\vspace{2pt}\noindent$\bullet$\textit{ Personal credit score analysis.} Here, we focus on another realistic workload which is more complex. Credit scoring is a typical application that needs to be protected more carefully - both the credit card holder's cost calendar and card issuer's scoring algorithm need to be remain secret. We implement a BP neural network-based credit scoring algorithm~\cite{jensen1992using} that calculates user's credit scores. The input file contains users' history records and the output is a {\em judgement} whether the bank should approve the next transaction. The model is trained on 10000 records and then is used to make prediction (i.e., output an output of confidence probability) on different number of test cases.

\begin{figure}[htbp]
\centerline{\includegraphics[scale=0.48]{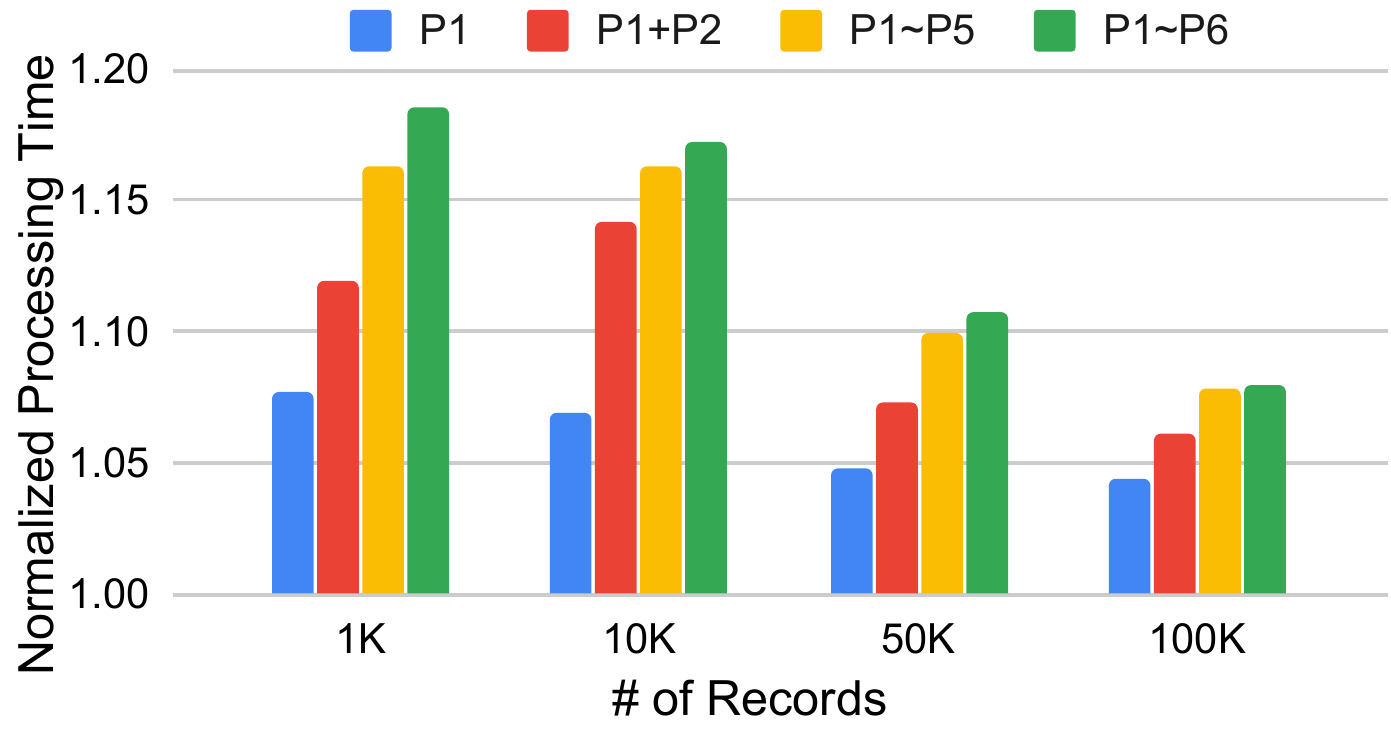}}
\caption{Performance on credit scoring}\label{fg-credit-score}
\end{figure}

As shown in Figure~\ref{fg-credit-score}, for testing on 1000 and 10000 records, P1\textasciitilde P5 would yield over 15\% overhead. 
While processing more than 50000 records, the overhead of the full check's performance does not exceed 25\%. And the overhead of P1\textasciitilde P6 does not exceed 10\% when processing 100K records.

\begin{figure}[htbp]
\centerline{\includegraphics[scale=0.48]{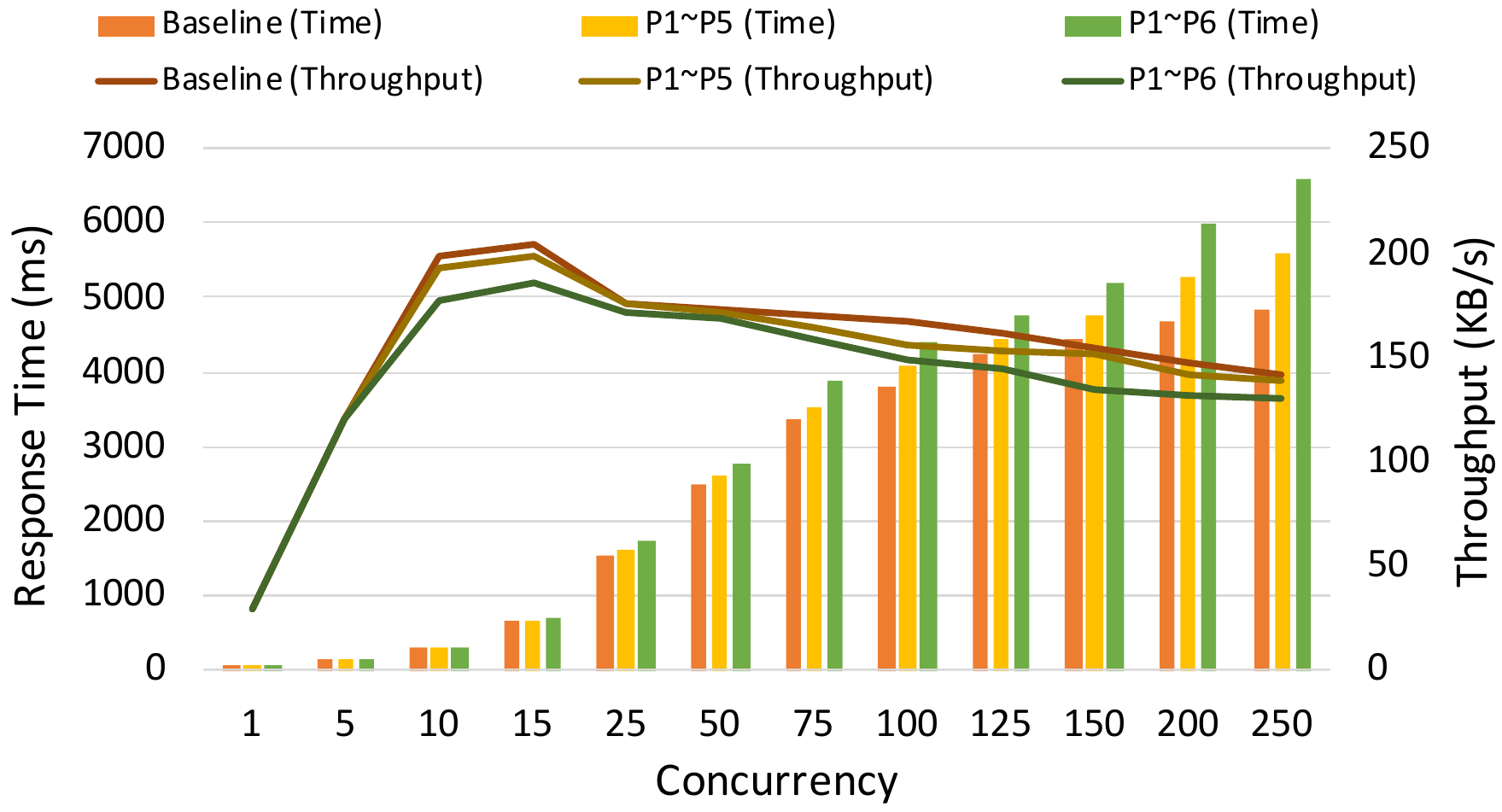}}
\caption{Performance on HTTPS server}\label{fg-https-all}
\end{figure}

\vspace{2pt}\noindent$\bullet$\textit{ HTTPS server.} 
To evaluate how our design would affect web services, we built an HTTPS server within an enclave using mbed TLS library~\cite{mbedtls}. And we only allow two system calls (\texttt{send/recv}) to be executed via the OCall stubs for client/server communication. A client executes a stress test tool - Siege~\cite{siege} - on another host in an isolated LAN. Siege is configured to send continuous HTTPS requests (with no delay between two consecutive ones) to the web server for 10 minutes. We make a comparison among the concurrent connections of pressure tests to see how our instrumented HTTPS server implementation would perform. 


Figure~\ref{fg-https-all} shows the response times and throughput for full checks being applied onto the SSL server benchmark. When the concurrency is less than 75, the instrumented HTTPS server has similar performance of the original in-enclave https server. When concurrency increases to 100, the response time and the throuphput drop to some extent. 
While after the concurrency increases to 150, the response time of instrumented server escalates significantly. On average, enforcing P1\textasciitilde P6 results in 14.1\% overhead on the response time. As for throughput, when the number of the concurrent connections is between 75 and 200, the baseline always achieves higher throughput, though just less than 10\% higher. All these experiments on realistic workloads show that proper policies on side channel mitigations can be enforced with only reasonable overhead.
}
\section{Discussion}\label{sec-discussion}

In previous sections we have shown that the design of CAT offers lightweight and efficient in-enclave verification of privacy policy compliance. 
Here we discuss some extensions.



\vspace{3pt}\noindent\textbf{Supporting other side/covert channel defenses}.
In Section~\ref{subsec-producer}, we talked about policy enforcement approaches for side channel resilience.
It demonstrated that our framework can take various side channel mitigation approaches to generate code carried with proof. Besides AEX based mitigations which we learnt from Hyperrace~\cite{chen2018racing}, others~\cite{doychev2015cacheaudit,almeida2016verifying,shih2017t,gruss2017strong,wu2018eliminating,wang2019identifying,orenbach2019cosmix} can also be transformed and incorporated into the CAT design. 
Even though new attacks have been kept being proposed and there is perhaps no definitive and practical
solutions to all side/covert channel attacks, we believe eventually some efforts can be integrated in our work. 

\vspace{3pt}\noindent\textbf{Supporting SGXv2}.
Our approach currently relies on SGXv1 instructions that prevents dynamically changing page permissions using a software DEP. The design could be simplified with SGXv2 instructions~\cite{mckeen2016intel} since dynamic memory management inside an enclave is allowed and protected in SGXv2 hardware. However, Intel has not shipped SGXv2 CPUs widely. So we implement the CAT model on SGXv1 to maximize its compatibility. 



\vspace{3pt}\noindent\textbf{Supporting multi-user}.
Currently we only support single user scenarios. Of course for multi-user scenarios, we can easily add a data cleansing policy which ensures that once the task for one data owner ends, all her data will be removed from the enclave before the next owner's data is loaded, together with the content of SSA and registers, while not destroying the bootstrap enclave after use.
Further, to fully support multi-user in-enclave services, we need to ensure each user's session key remains secret and conduct remote attestation for every user when they switch. Hardware features like Intel MPX~\cite{shen2018isolate} can be applied to enforce memory permission integrity~\cite{zhao2020mptee}, as a supplementary boundary checking mechanism.

\vspace{3pt}\noindent\textbf{Supporting multi-threading}.
When taking multi-threading into account, the proof generation process become more complicated and cumbersome~\cite{guo2007certified}. Furthermore, multi-threading would introduce serious bugs~\cite{weichbrodt2016asyncshock}. However, auditing memory read operations from other threads seems taking the multi-threading leakage once and for all. Actually, if we don't prevent attacks mentioned in CONFirm~\cite{xu2019confirm}, the proof enforcement of CFI is still broken due to a time of check to time of use (TOCTTOU) problem. To cope with that, we can make all CFI metadata to be read from a register instead of the stack, and guarantee that the instrumented proof could not be written by any threads~\cite{burow2019sok}. 




\vspace{3pt}\noindent\textbf{Supporting on-demand policies}.\label{subsec:morescenario}
The framework of our system is highly flexible, which means assembling new policies into current design can be very straightforward. Different on-demand policies can be appended/withdrawn to serve various goals. For example, we can attach additional instrumentation to the code and the policy enforcement to the in-enclave verifier in case of the discovery of new side/covert channels and newly-published security flaws. CAT can make the quick patch possible on software level, just like the way people coping with 1-day vulnerabilities - emergency quick fix. 
Besides, users can also customize the policy according to their need, e.g., to verify code logic and its functionalities.
\section{Related Work}\label{sec-relatedwork}

\vspace{3pt}\noindent\textbf{Secure computing using SGX}.
Many existing works propose using SGX to secure cloud computing systems, e.g., VC3~\cite{schuster2015vc3}, TVM~\cite{hynes2018efficient}, by using 
sand-boxing~\cite{hunt2018ryoan}, containers~\cite{arnautov2016scone}, and library OSes~\cite{tsai2017graphene,shen2020occlum}. 
These systems relies on remote attestation to verify the platform and the enclave code, as a result, they either do not protect the code privacy or they consider a one-party scenario, i.e., the code and data needed for the computation are from the same participant. In contrast, we consider 3 real world scenarios (CCaaS, CDaaS and CDCM) protecting code and data from multiple distrustful parties. 

\vspace{3pt}\noindent\textbf{Data confinement with SFI}.
Most related to our work are data confinement technologies, which confines untrusted code with confidentiality and integrity guarantees. Ryoan~\cite{hunt2018ryoan} and its follow-up work~\cite{hunt2018chiron} provide an SFI-based distributed sand-box by porting NaCl to the enclave environment, confining untrusted data-processing modules to prevent leakage of the user’s input data. 
However the overhead of Ryoan turns out huge (e.g., 100\% on genes data) and was evaluated on an software emulator for supporting SGXv2 instructions.
XFI~\cite{erlingsson2006xfi} is the most representative unconventional PCC work based on SFI, which places a verifier at OS level, instead of a lightweight TEE. Occlum~\cite{shen2020occlum} is 
a design of SGX-based library OS, enforcing in-enclave task isolation with  MPX-based multi-domain SFI scheme. 
As the goal of SFI scheme is not to prevent information leakage from untrusted code, none of them employs protections against side channel leakages.

\vspace{3pt}\noindent\textbf{Code privacy}.
Code secrecy is is an easy to be ignored but very important issue~\cite{mazmudar2019mitigator,kuccuk2019managing}.
DynSGX~\cite{silva2017dynsgx} and SGXElide~\cite{bauman2018sgxelide} both make possible that developers execute their code privately in public cloud environments, enabling developers to better manage the scarce memory resources. However, they only care about the developer's privacy but ignore the confidentiality of data belonging to users. 

\vspace{3pt}\noindent\textbf{Confidentiality verification of enclave programs}.
With formal verification tools, Moat~\cite{sinha2015moat} and its follow-up works~\cite{sinha2016design} can verify if an enclave program has the risk of data leakage. The major focus of them is to verify the confidentiality of an SGX application outside the enclave formally and independently. Although it is possible that the verification could be performed within a ``bootstrap enclave'', the TCB would include the IR level language (BoogiePL) interpreter~\cite{barnett2005boogie} and a theorem prover~\cite{de2008z3}. Moreover, neither of them can discharge the large overhead introduced by instruction modeling and assertion proving when large-scale real-world programs are verified.



\vspace{3pt}\noindent\textbf{Side channel attacks and defenses}.
Side channels pose serious threats to secure computing using SGX as attackers can use them to circumvent explicit security defenses implemented by SGX. A rich literature has focused on discovering SGX side channels~\cite{lee2017inferring,wang2017leaky,van2018foreshadow,chen2019sgxpectre} and their defenses~\cite{shinde2016preventing,shih2017t,oleksenko2018varys,sinha2017compiler,chen2018racing}. Existing SGX secure computing work often assumes side channels as an orthogonal research topic~\cite{sinha2015moat,subramanyan2017formal,shen2020occlum}. 
Our framework is designed with side channels in mind and we have shown that it can flexibly support integration of instrumentation based side channel defenses.

\section{Conclusion}\label{sec-conclusion}

In this paper we proposed the CAT, a remote attestation model that allows the user to verify the code and data provided by untrusted parties without undermining their privacy and integrity. Meanwhile, we instantiated the design of a code generator and a code consumer (the bootstrap enclave) - a lightweight PCC-type verification framework.
Due to the differences between normal binary and SGX binary, we retrofit the PCC-type framework to be fitted into SGX. In return, we reduce the framework's TCB as small as possible.
Our work does not use formal certificate to validate the loaded private binary, but leverage data/control flow analysis to fulfill the goal of verifying if a binary has such data leakage, allowing our solution to scale to real-world software. Moreover, our method provides a new paradigm for PCC to use a TEE (other than the OS) as an execution environment, which provides more powerful protection. 
 


\normalem

\bibliographystyle{./bibliography/ACM-Reference-Format}
\bibliography{./bibliography/ms}

\appendix

\section*{Appendix}

\subsection{Instrumentation Details}\label{appendix-instrumentation}

Here we illustrate other instrumentation modules in our code generator.

\vspace{3pt}\noindent\textbf{RSP modification instrumentation}. Since RSP spilling would cause illegal implicit memory writing, RSP modification instructions should also be checked. This module first locates all RSP modification instructions in the program and then instruments assembly code after them to check whether the RSP values are out of bounds. Just like storing instruction instrumentation, the upper and lower boundaries of RSP are specified by the loader and written into the assembly instructions by the rewriter, while the compiler only fills them with speical immediates (5ffffffffffff and 6ffffffffffff).

When the instrumentation finds that the stack pointer is modified to an illegal address, it will cause the program to exit. Fig.~\ref{fg-rsp} shows eight instructions be inserted after the \texttt{ANDQ} instruction, which is tend to reserve new stack spaces (minus 16 from the value in RSP register). We leave the enforcement of implicit modification of the stack pointer using \texttt{PUSH} and \texttt{POP} by adding guard pages (a page with no permission granted) to the dynamic loader.

\begin{figure}[H]
\begin{center}
\begin{minipage}{0.3\textwidth}
\begin{lstlisting}[basicstyle=\footnotesize,numbers=left]
andq    $-16, %rsp
pushq   %rax
movabsq $0x5FFFFFFFFFFFFFFF, %rax
cmpq    %rax, %rsp
ja      exit_label
movabsq $0x6FFFFFFFFFFFFFFF, %rax
cmpq    %rax, %rsp
jb      exit_label
popq    %rax
\end{lstlisting}
\end{minipage}
\end{center}
\caption{RSP Modifying Instrumentation}\label{fg-rsp}
\end{figure}

\vspace{3pt}\noindent\textbf{Indirect branch instrumentation}. For checking indirect branches, we first extract all legal target names at assembly level, and output them to a list. 
After that, we insert a inspection function calling in front of every indirect branch instruction (in Fig.~\ref{fg-indirect}), to achieve forward-edge CFI check at runtime. 
Specifically, the inspection function \verb|CFICheck| is written and included in the target binary, to search if the indirect branch is on that list, therefore ensuring they conform to the program control flow. 

\begin{figure}[hbp]
\begin{center}
\begin{minipage}{0.3\textwidth}
\begin{lstlisting}[basicstyle=\footnotesize,numbers=left]
movq	%reg, %rdi
callq	CFICheck
\end{lstlisting}
\small{Instrumentations before callq \*\%reg}
\end{minipage}
\begin{minipage}{0.3\textwidth}
\begin{lstlisting}[basicstyle=\footnotesize,numbers=left]
movq	(%reg), %rdi
callq	CFICheck
\end{lstlisting}
\small{Instrumentations before callq \*(\%reg)}
\end{minipage}
\end{center}
\caption{Indirect Call Instrumentation}\label{fg-indirect}
\end{figure}

\vspace{3pt}\noindent\textbf{Shadow stack}. For function returns, the code generator instruments instructions to support a shadow call stack, which is a fully precise mechanism for protecting backwards edges~\cite{burow2019sok}. The shadow stack’s base address is specified by the loader, and will be rewritten by the Imm rewriter (to replace the imm filled in by the compiler in advance).

As shown in Fig.~\ref{fg-shadowstack}, at every function entry, we insert instructions (before the function stack alignment) that will modify the shadow stack top pointer and push the function's return address into the shadow stack. Similar to instrumentation at the function entry, instructions inserted before the function returns  modify the stack pointer and pop the return address. Comparing the saved return address with the real return address, \texttt{RET} will be checked.

\begin{figure}[hbp]

\begin{minipage}{0.25\textwidth}
\begin{center}
\begin{lstlisting}[basicstyle=\footnotesize,numbers=left]
movabsq	$0x2FFFFFFFFFFF, %r11
addq	$8, (%r11)
movq	(%r11), %r10
addq	%r10, %r11
movq	(%rsp), %r10
movq	%r10, (%r11)
pushq	%rbp
movq	%rsp, %rbp
\end{lstlisting}
\small{Instrumentation before stack alignment}
\end{center}
\end{minipage}
\hfill
\begin{minipage}{0.3\textwidth}
\begin{center}
\begin{lstlisting}[basicstyle=\footnotesize,numbers=left]
movabsq	$0x2FFFFFFFFFFF, %r11
movq	(%r11), %r10
addq	%r11, %r10
subq	$8, (%r11)
movq	(%r10), %r11
cmpq	%r11, (%rsp)
jne	exit_label
\end{lstlisting}
\small{Instrumentation before function return}
\end{center}
\end{minipage}
\hfill
\begin{minipage}{0.3\textwidth}
\begin{center}
\begin{lstlisting}[basicstyle=\footnotesize,numbers=left]
exit_label:
    movl    $0xFFFFFFFF, %edi
    callq   exit
\end{lstlisting}
\small{Instrumentation for exit label}
\end{center}
\end{minipage}

\caption{Structured Guard Formats of Shadow Stack}\label{fg-shadowstack}

\end{figure}

\vspace{3pt}\noindent\textbf{SSA monitoring instrumentation}. 
As demonstrated in previous works~\cite{gruss2017strong,chen2018racing}, AEX can be detected by monitoring the SSA. Therefor, to enforce P6, we instrument every basic block to set a marker in the SSA and monitor whether the marker is overwritten by AEX within the basic block. The execution is terminated once the number of AEXes within the basic block exceeds a preset threshold. 

A function is also implemented to get the interrupt context information in the bootstrap enclave's SSA area. At the beginning of each basic block, we call this function through instrumentation to check whether there are too many interruptions during execution. When a basic block is too large, this function will also be called in the middle of basic block every k ($k=20$) instructions. We count the number of interrupts/AEXs that occurred from the last check to the current check. When 22 or more are triggered, the target program aborts.

\vspace{3pt}\noindent\textbf{Alternatives}. 
To mitigate AEX based side-channel risks, CAT-SGX provides an alternative enforcement mechanisms, through TSX, which can be chosen when compiling the target program. The TSX approach is based upon T-SGX~\cite{shih2017t}, putting transaction memory protection on each basic block and running a fall-back function to keep track of the number of interrupts observed. Just like T-SGX, when more than 10 consecutive AEXes happen, the computation aborts, due to the concern of an ongoing side-channel attack.  The protection is instrumented by the generator and its presence is verified by the code consumer in the enclave. 

We have implemented a function, in which \texttt{XBEGIN} and \texttt{XEND} is called and fallback is specified. Around each branch and \texttt{CALL}/\texttt{RET} instruction and at the begin/end of each basic block, we call this function so that the program leaves the last transaction and enters a new transaction when a possible control flow branch occurs and completes. Some code snippets are shown in Figure~\ref{fg-tsgx}. 

To deal with the compatibility problems caused by calling functions that has no need to be checked (e.g., the system calls via OCall stubs), we implemented another non-TSX wrapper for external functions.
For instance, our pass will generate an alternative function \verb|wrapper_foo| to replace original function \verb|foo|, to avoid the TSX instrumentation.

\begin{figure}
\begin{center}




\begin{minipage}{0.3\textwidth}
\begin{lstlisting}[basicstyle=\footnotesize,numbers=left]
movq	%rax, %r15
lahf
movq	%rax, %r14
callq	transactionEndBegin
movq	%r14, %rax
sahf
movq	%r15, %rax
\end{lstlisting}
\end{minipage}

\end{center}
\caption{TSX instrumentation}\label{fg-tsgx}
\end{figure}

\subsection{Preparing Target Binary}\label{appendix-preparing}

\vspace{3pt}\noindent\textbf{Libc}. To manage interactions with the untrusted operating system, we make some Ocall stubs for system calls. 
Related works~\cite{shinde2017panoply,tsai2017graphene,priebe2019sgx,shindebesfs} provide various great Ocall interfaces. But some of them still require additional interface sanitizations.
We use parts of Musl Libc~\cite{musllibc} for completing the code loading support (Section~\ref{subsec:code-loading-support}).
Undoubtedly, the Musl Libc also should be instrumented. Then, it can be linked against other necessary libraries statically, e.g., mbedTLS for buiding an HTTPS server.

\vspace{3pt}\noindent\textbf{Stack and heap}. We also reserved customized stack and heap space for the target program execution. During the above-mentioned loading phase, the CAT-SGX system will initialize a 4MB size memory space for the stack, and will link against a customized and instrumented \verb|malloc| function for later heap usage. In current version of our prototype, the memory ranges of the additional stack and the heap provided for the target program are fixed, for efficient boundary checking.

\vspace{3pt}\noindent\textbf{Other necessary functions}. The instrumented proof includes not only the assembly instructions. Some necessary functions and objects also should be compiled and linked.
Since we need an algorithm to check if the address of an indirect branch target is on the legal entry label list (for P5 enforcement), a binary search function \verb|CFICheck| is inserted into the target program.
Similarly, as we need a function to enforce P6, necessary functions need to be called for SSA monitoring frequently. 
Those objects would also be disassembled and checked during the stage of proof verification, to ensure that these can not be compromised when they are called.


\end{document}